\documentclass[10pt,journal,compsoc]{IEEEtran}
\usepackage{array}
\usepackage{verbatim}
\usepackage{multirow}
\usepackage{graphicx}
\usepackage[unicode=true]{hyperref}
\usepackage{multicol}% http://ctan.org/pkg/multicols
\usepackage[export]{adjustbox}
\usepackage{enumitem}
\usepackage{algorithm,algorithmicx}
\usepackage[noend]{algpseudocode}
\ifCLASSOPTIONcompsoc
\usepackage[nocompress]{cite}
\else
\usepackage{cite}
\fi

\newsavebox{\mybox}

\begin{document}

\title{Protecting Military Avionics Platforms from Attacks on MIL-STD-1553 Communication Bus}

\author{Orly~Stan,~Yuval~Elovici,~Asaf~Shabtai,~Gaby~Shugol,~Raz~Tikochinski,~Shachar~Kur% <-this % stops a space
	\IEEEcompsocitemizethanks{\IEEEcompsocthanksitem O. Stan, Y. Elovici, and A. Shabtai are with the Department of Software and Information Systems Engineering, Ben-Gurion University of the Negev.\protect\\
		% note need leading \protect in front of \\ to get a newline within \thanks as
		% \\ is fragile and will error, could use \hfil\break instead.
		E-mails: stan@post.bgu.ac.il, \{elovici,shabtaia\}@bgu.ac.il
		\IEEEcompsocthanksitem G. Shugol, R. Tikochinski, and S. Kur are with Astronautics C.A. ltd. \protect\\
		E-mails: \{g.shugol, r.tikochinski, s.kur\}@astro.co.il}}% <-this % stops an unwanted space}

\IEEEtitleabstractindextext{
	\begin{abstract}
		MIL-STD-1553 is a military standard that defines the physical and logical layers, and a command/response time division multiplexing of a communication bus used in military and aerospace avionic 
		platforms for more than 40 years. As a legacy platform, MIL-STD-1553 was designed for high level of fault tolerance while less attention was taken with regard to security. Recent studies already addressed the impact of successful cyber attacks on aerospace vehicles that are implementing MIL-STD-1553. 
		In this study we present a security analysis of MIL-STD-1553.
		In addition, we present a method for anomaly detection in MIL-STD-1553 communication bus and its performance in the presence of several attack scenarios implemented in a testbed, as well as results on real system data. Moreover, we propose a general approach towards an intrusion detection system (IDS) for a MIL-STD-1553 communication bus.
	\end{abstract}
	
	}

\maketitle
\IEEEdisplaynontitleabstractindextext
\IEEEpeerreviewmaketitle

\section{Introduction}

\IEEEPARstart{M}{IL-STD-1553} is a military standard developed by the US Department of Defense (DoD) for the purpose of military platform integration \cite{editionmil} which has served as the backbone of military and aerospace avionic platforms (e.g., F-15, AH-64 Apache, F-16, V-22, X-45A, F-35)
for more than 40 years. It is primarily used for mission-critical systems that require a high level of fault tolerance, since it is deterministic and dual redundant; it also uses a reduced cable topology, connecting all devices on a single bus in a multipoint topology, as opposed to point-to-point topologies.

MIL-STD-1553 is considered deterministic, because it is based on a master/slave methodology in which the master issues messages based on a predefined order and timing. 
Although other modern, reliable and deterministic data buses have been introduced \cite{deshu1991guilin,gillen1992introduction}, MIL-STD-1553 remains the most widely used standard in military aviation as it has been for the last 40 years, and is expected to be used in the future. 
The main reason that alternative deterministic communication buses are not used in existing platforms is the difficulty of modifying an entire operational platform and replacing the main data transmission topology. Moreover, subsequent standards are based on the communication protocol defined by MIL-STD-1553.
For these reasons, MIL-STD-1553 will likely be an integral component of critical military platforms for many more years to come.

MIL-STD-1553 was developed long before the notion of cyber security was  familiar and even basic cyber attacks, such as denial-of-service (DoS) attacks \cite{gligor1983note}, had not yet been introduced. 
Research regarding DoS attacks initially reported in the early 1980s, several years after the release of the most recent version of MIL-STD-1553 in 1978, and focused mainly on DoS in operating systems, rather than computer networks \cite{gligor1983note}. The Designer's Notes for MIL-STD-1553 include a chapter discussing several aspects of network system security which should be addressed when implementing a 1553 communication bus \cite{editionmil}:

\begin{itemize}[noitemsep,nolistsep]
	\item system security policy  -- defines the classification levels of the system, data, and personnel that are related to the communication bus;
	\item system security architecture -- specifies four approaches for designing systems that process classified plain text data and unclassified data;
	\item Tempest -- states that all components processing unencrypted classified data should be protected against compromising emanation;
	\item Encryption -- should be used in order to isolate components with different classification levels from classified data;
	\item Trusted message routing and control design -- maintaining low bit error rate, parity coding of control words, and monitoring the bus controller can help in detecting errors in messages or in their routing.
\end{itemize}

Although the Designer's Notes provide references to security aspects, they only contain general guidelines, including references to standards that might not be appropriate for all MIL-STD-1553-based systems (e.g., military vessels developed by other countries might have different or additional compliance requirements than those defined in MIL-STD-1553). Moreover, because the standard is defined for military purposes, more specific guidelines cannot be provided due to confidentiality requirements. Finally, the standard is implemented by various types of systems with diverse objectives, which makes it extremely complicated to provide more specific requirements will suite all existing systems.

Therefore, despite the attention paid to security issues in the Designer's Notes, MIL-STD-1553 still contains vulnerabilities that expose the platforms implementing it to cyber attacks, which are not addressed in the Designer's Notes or in the various updates that have been made to the standard since its first release. Section \ref{sec:Security-Analysis-of} provides a security analysis of MIL-STD-1553 and discusses possible attack methods and their consequences. 

As cyber attacks play a major role in modern warfare and since military platforms are likely to be attractive targets for attackers \cite{miller2012scada},\cite{lindsay2013stuxnet}
, it has become clear that the systems implementing the MIL-STD-1553 standard require improved protection.
Due to its widespread deployment in many platforms, applying changes to the various components of the 1553 communication bus is cost prohibitive.
Hence, instead of securing the standard itself, we introduce a method for the detection of anomalous traffic transmitted over the 1553 communication bus.

Recent studies have addressed the impact of successful cyber-attacks on aerospace vehicles that implement MIL-STD-1553 
\cite{mcgraw2014cyber,nguyen2015towards}. In \cite{nguyen2015towards} the author presents some of the associated vulnerabilities and suggests theoretical methods for creating covert channels over the communication bus. The authors in \cite{mcgraw2014cyber} illustrated the physical impact of simulated cyber-attack on an aerospace vehicle. However, none of them proposed a solution for detect and/or prevent such attacks.

In this paper we present a security analysis of the MIL-STD-1553 communication protocol and propose a supervised sequence-based anomaly detection method for identifying cyber-attacks. For evaluating our method we established an operational testbed in which we executed three attack scenarios. All three attacks were perfectly identified by our proposed method. Moreover we evaluated the learning process of the proposed method on datasets collected from a real system; this evaluation indicates that a very short period of time (two to five seconds) is sufficient for achieving a very low false alarm rate.

The following sections provide: an overview of the MIL-STD-1553 architecture and communication protocol (Section \ref{sec:1553-background}); a review of related works regarding the security of systems implementing MIL-STD-1553 and other communication bus technologies, (Section \ref{sec:related-works}); a security analysis which defines the assets of a MIL-STD-1553 bus, the attacker profile, and possible threats to the communication bus (Section \ref{sec:Security-Analysis-of}); 
proposal for a sequence-based anomaly detection method for a MIL-STD-1553 communication bus (Section \ref{sec:seq-timing-module}); 
description of the testbed established for evaluating the proposed method and evaluation results for both simulated and real-system scenarios (Section \ref{sec:experiment-and-results}); finally Section \ref{sec:discussion} discusses the performance and limitations of the proposed sequence-base anomaly detection method and suggests extensions to the method; Section \ref{sec:future-work} concludes the paper and presents future research direction.

\begin{figure}[h!]
	\centering
	\includegraphics[scale=0.26]{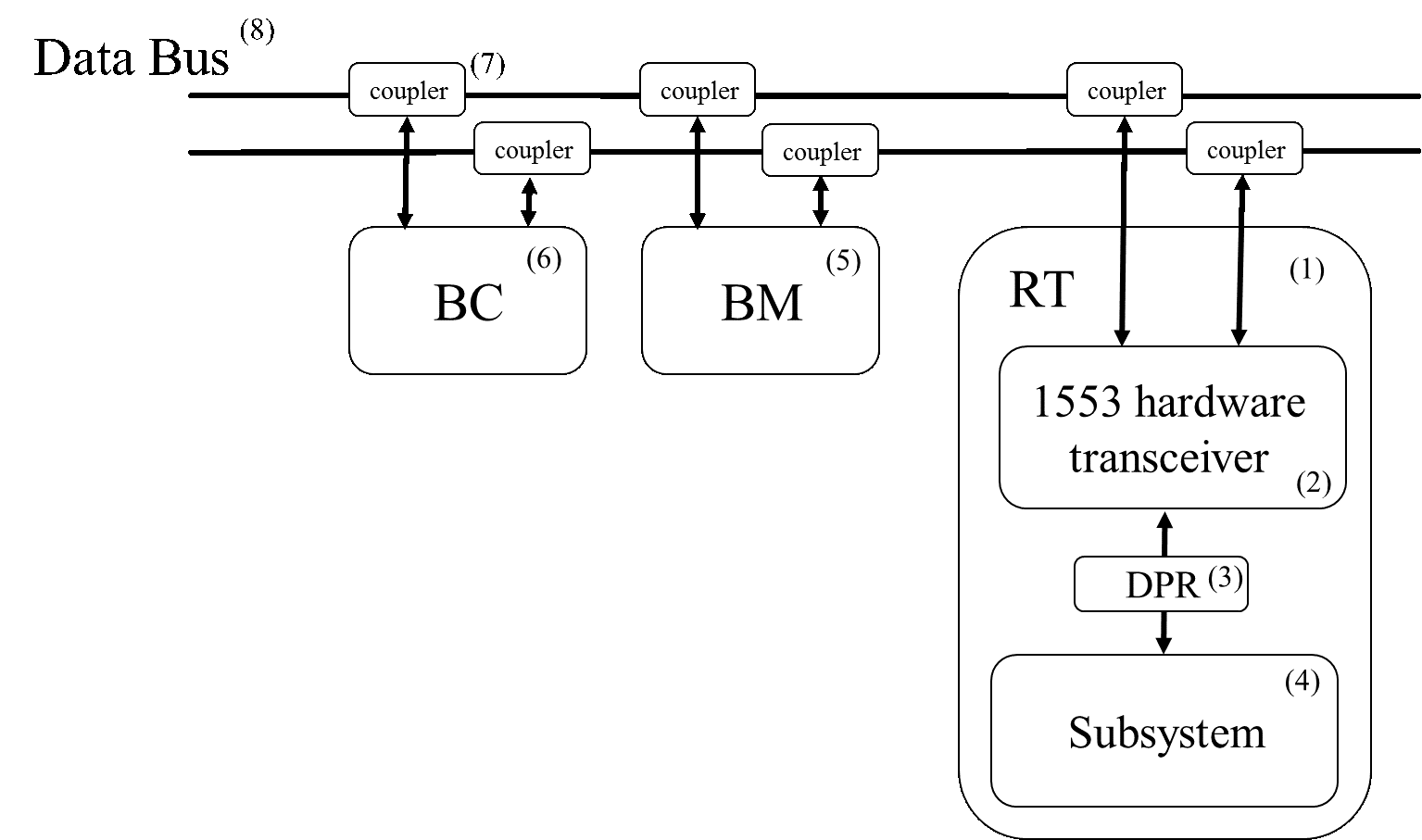}
	\caption{The MIL-STD-1553 bus architecture and its primary components.}
	\label{fig:MIL-STD-1553-bus-architecture}
\end{figure}

\section{\label{sec:1553-background}1553 Communication Bus - Background}

MIL-STD-1553 defines a dual redundant serial communication bus used for transmitting data between a bus controller and remote terminals using a multipoint, master-slave bus topology. It was first published in 1973 and the latest version, MIL-STD-1553B, published in 1978, is still used in many military and aerospace systems to this day. MIL-STD-1553 defines the physical layer of the communication bus as well as the logical layer and a command/response time division multiplexing methodology using a 1Mbps transfer rate data bus, while specifying the transmission timings.

\subsection{Bus architecture}

The 1553 communication bus includes five key elements: Remote terminal (RT), bus controller (BC), bus monitor (BM), coupler, and the bus itself (illustrated in Figure \ref{fig:MIL-STD-1553-bus-architecture}). The bus is redundant -- if a message cannot be transmitted on the main channel it will be retransmitted on the backup channel. Although there are redundant channels, only one element can transmit data over the bus at a time.
All elements connected to the bus are continuously exposed to the data transmitted, even if not designated for them. The communication is managed by the BC, and all other elements follow its commands. The bus can support up to 31 connected remote terminals.

\textbf{Remote~terminal~(RT).} Consists of three components. The hardware transceiver is responsible for data transfer between the bus and the subsystem. It is connected directly to the bus and exchanges data with the subsystem via a dual port RAM (DPR). In addition, it must be able to decode and buffer messages, detect transmission errors, and perform data validation tests. Invalid data should be discarded. The DPR is shared memory which enables data transfer between the transceiver and the subsystem. Both the transceiver and subsystem have read and write permission to this memory. The subsystem is the computational unit (platform computer) of the RT. The subsystem is responsible for all data processing and calculations required for the system to function.

\textbf{Bus~controller~(BC).} Responsible for managing the communication between the RTs connected to the bus using command/response messages.
It is the only component that initiates data transfers on the bus to/from RTs or between two RTs. There may be several terminals with BC capabilities connected to the same bus for backup, but only one of them can function as the active BC at a given time. The BC initiates commands to the RTs based on a predefined order and timing.

\textbf {Bus~monitor~(BM).} Responsible for listening and collecting data from the bus in order to observe the state and operational mode of the system and subsystems. The BM is a passive device and does not send any messages, and therefore cannot provide a status report on the data transferred over the bus.

\textbf{Coupler.} A physical component used to isolate the components connected to the bus from one another and eliminate the possibility of damage to the bus in case one of the components malfunctions.

\textbf{Data~Bus.} The transmission medium that physically enables all communication between the components connected to it.
	
\subsection{\label{subsec:communication-protocol}Communication protocol}

Words are the data structure used for transmitting commands, data, and status over the bus. A collection of words defines a message used for receiving or transmitting data. Messages can be periodic or aperiodic. Periodic messages are sent at fixed time intervals (i.e., time cycles). A major frame is a predefined time frame in which all periodic messages are transmitted at least once (derived from the periodic message with the longest time cycle). Aperiodic messages are event-driven and therefore are not sent in fixed time cycles. However, they have a fixed time slot in the major frame.

\begin{figure}[t]
	\centering
	\includegraphics[scale=0.37]{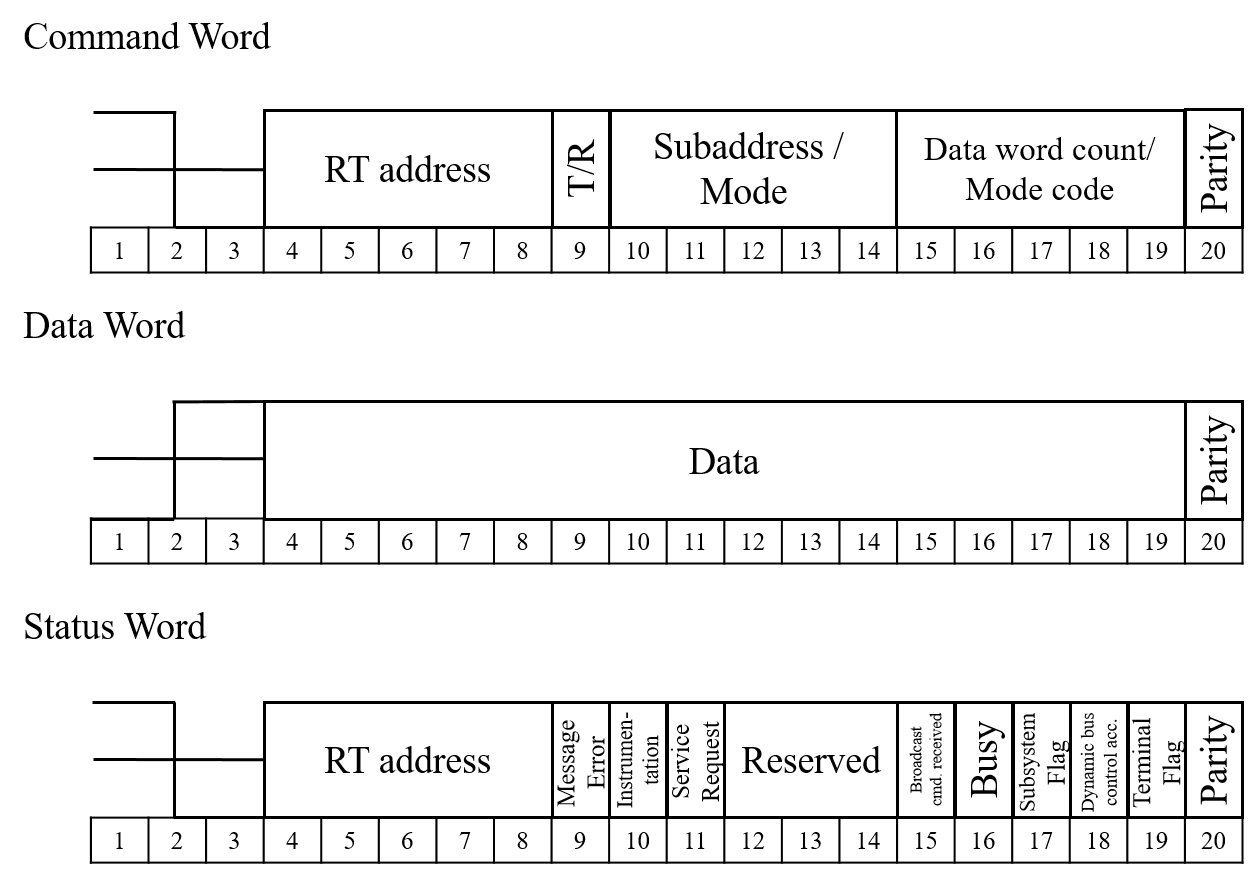}
	\caption{Communication protocol words structure.}
	\label{fig:Communication-protocol-words}
\end{figure}

The standard defines three types of words: command, data, and status  (illustrated in Figure \ref{fig:Communication-protocol-words}). All words are 20 bits long, starting with three bits of synchronization and ending with a parity bit.

\begin{description}
\item [{Command~word.}] Initiated by the BC and designated to an RT. The command specifies the action that the RT should perform: whether to receive or transmit data. The remaining 16 bits are defined as follows:

	\emph{Terminal address (TA)} -- a five bit field containing the address of the RT that the command is designated for. It can contain up to 31 RT addresses (00000B to 11110B), since the terminal address 11111B is reserved for broadcast command.
	
	\emph{T/R bit} -- a single bit that indicates the direction of the required data transfer. Logic 1 indicates that the RT should transmit data, and logic 0 indicates that the RT should receive data.

	\emph{Subaddress/Mode} -- a five bit field indicating the subaddress of the RT to receive/transmit the data, or that this command is a mode code (in this case it is set to 00000B or 11111B). 
	Mode codes are special commands used to change the operation mode of the RTs such as: timing synchronization, RT transmitter shut down, and request to initiate self-test.
	
	\emph{Data word count/mode code} -- a five bit field which contains the number of data words to be received/transmitted. If a mode code is set, these five bits indicate the mode code.

\item [{Data~word.}] Contains the actual data being transferred on the bus. There is no predefined structure for data words.

\item [{Status~word.}] Sent by the RT to the BC upon receiving a valid message, in order to report its status to the BC. It contains different flags indicating different types of errors, such as received data error, data processing error, and circuitry error. It also allows the RT to request a service from the BC. 
\end{description}

\begin{figure}[h]
	\centering
	\includegraphics[scale=0.3,keepaspectratio]{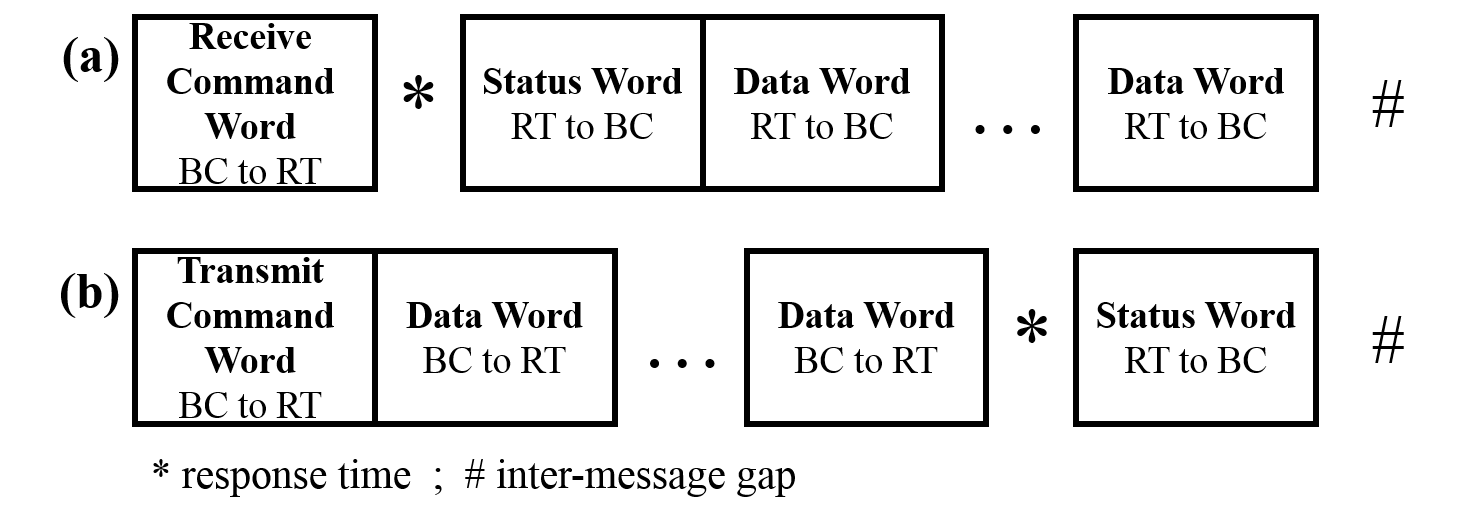}
	\caption{BC-RT (a) and RT-BC (b) transfer format.}
	\label{fig:BC-RT-and-RT-BC}
\end{figure}

\begin{figure}[h]
	\centering
	\includegraphics[scale=0.30,keepaspectratio]{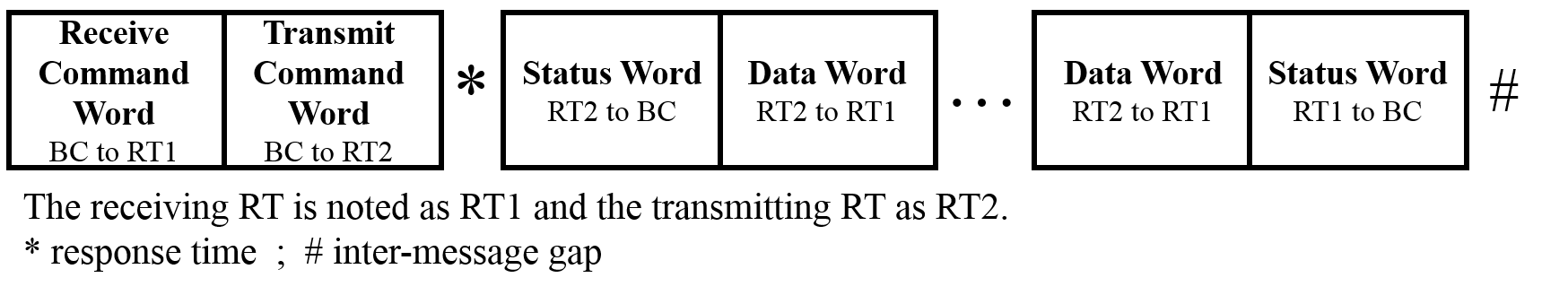}
	\vspace{-\baselineskip}
	\caption{RT-RT transfer format.}
	\label{fig:RT-RT-transfer-format}
\end{figure}

\subsection{\label{subsec:Communication-formats}Communication formats}

There are four types of communication between elements over the bus, all of which are initiated by the BC. The communication formats are designed to maintain high reliability of the protocol by acknowledging every message sent on the bus and flagging for errors and validation of the messages (using status words).

\textbf{BC-RT/RT-BC~data~transfer.} The communication between the BC and an RT has two formats: 'receive' (BC-RT) and 'transmit' (RT-BC). In order to initiate a BC-RT communication (Figure \ref{fig:BC-RT-and-RT-BC}(a)), the BC issues a 'receive' command to the RT, and immediately transmits the data words. The RT receives and validates the data, and responds with a status word. In order to start a RT-BC communication (Figure \ref{fig:BC-RT-and-RT-BC}(b)), the BC issues a 'transmit' command to the RT. The RT receives the command and responds with a status word, which is immediately followed by the data words it should transmit.

\textbf{RT-RT~data~transfer.} In RT-RT communication (Figure \ref{fig:RT-RT-transfer-format}), one RT transmits data
	to another RT. The BC starts the communication by issuing a 'receive'
	command to the receiving RT, which is immediately followed by sending a 'transmit' command to the transmitting RT. The transmitting RT responds with a status word and transmits its data words. Upon receiving the data,
	the receiving RT responds with a status word.

\textbf{Mode~code~transfer.} The BC can send a mode command by setting the subaddress/mode field to 00000B or 11111B. In this case the word count field defines which mode code should be performed. A mode command can be sent to a specific RT or to all RTs. A mode command can be associated with up to one data word.

\textbf{Broadcast~transfer.} The standard also supports broadcast messages. Broadcast can be used with messages in which only the BC is transmitting data and all others are receiving. The broadcast message format is similar to the non-broadcast messages, with two exceptions: the terminal address field is set to 11111B, and all receiving RTs suppress their status word transmission.

\section{\label{sec:related-works}Related Work}

Although MIL-STD-1553 is the basis for many mission-critical platforms, there has been very little research conducted regarding its security. 
The security of mission-critical and embedded systems was discussed in \cite{chong2005Survivability,jiang2010periofic,vai2016design}.
In 2005, Chong et al. \cite{chong2005Survivability} suggested design principles and guidelines for a survivability system architecture and applied it to a DoD information system.
In 2016, Vai et al. \cite{vai2016design} developed a methodology for designing a general mission-critical embedded system that considers cyber security aspects. The authors suggest a modular system architecture that contains cyber security features (e.g., cryptographic components and a separation kernel), and monitoring and recovering services.

These kind of security measures are suitable for systems that are in their design phase, in which different security features and principles can be considered and integrated correctly; however, they are not suitable for enhancing the security of existing 1553 bus implementations, because changing components of the 1553 communication bus is cost prohibitive due to its extensive deployment in wide range of aircrafts and vehicles.

In the context MIL-STD-1553, in 2014, McGraw, et al. \cite{mcgraw2014cyber} explored the impact of malicious actions on a satellite that uses a 1553 communication bus for intercommunication between its subsystems. The communication bus was modeled using SimPy (a simulation framework written in Python), and consists of a BC, BM, and 10 RTs. In addition, STK SOLIS (a simulation environment for spacecraft) was used for generating a high fidelity model and data exchange between the simulated subsystems.
In order to explore malicious actions, McGraw, et al. \cite{mcgraw2014cyber} characterized the normal behavior of a space asset and used it to detect perturbations which may indicate the presence of a malware.
Two scenarios of abnormal behavior were simulated: the presence of solar flares or ionization activity, and the presence of a malware. These abnormal scenarios were simulated by injecting 
noises (of different magnitudes) into the sensors' models. 
Manual analysis of the results indicated that it is possible to detect the anomalous events. Moreover, the authors were able to distinguish between events that might be caused by ionization and those that might be caused by a malware. The authors also observed a significant change in the satellite physical position in the presence of malware.
The authors reached their conclusions by manually exploring the simulation results and did not provide an automatic method for detecting anomalies.

In contrast to the manual analysis performed in \cite{mcgraw2014cyber}, we propose a real-time machine learning-based IDS for detecting abnormal behavior. 
Furthermore, in order to evaluate our proposed solution we used an operational testbed composed of an actual 1553 communication bus (presented in more detail in Section \ref{subsec:Emulator-Architecture}) on which we were able to execute attacks that exploit various vulnerabilities in the protocol rather than data manipulation. The use of actual hardware enabled us to simulate data transmission in a more realistic environment and extract more precise features.

In 2015, Nguyen \cite{nguyen2015towards} introduced several methods for creating covert channels over a 1553 communication bus, in order to leak data from high security level subsystems to lower subsystems.
The suggested attacks utilize different features and behavioral characteristics of the communication protocol defined by the standard, in order to establish a signaling mechanism between two cooperating subsystems connected to the same communication bus.
Nguyen presented three attack scenarios and categorized them into two types: timing and storage attacks.
Timing attacks utilize time delays between messages defined by MIL-STD-1553, while storage attacks utilize word structure and programmer-defined features. More specifically, the storage attacks utilize the 'command illegalization' implementation (which is a programmer defined feature), and the Service Request (SR) feature defined by the standard which enables an RT to notify the BC that it needs to transmit or receive data.
The suggested attack scenarios are merely theoretical and were not empirically tested. In addition, the suggested attacks rely on assumptions which are not necessarily correct or applicable for all 1533 communication buses. Moreover, the attacks presented are inefficient. For example, if the RT that executes a timing attack is able to control its response delays to the granularity of one microsecond, it can leak up to three bits per message.

Covert channel attacks were taken into consideration in the security analysis we present in Section \ref{sec:Security-Analysis-of} and in our proposed solution described in Section \ref{sec:seq-timing-module}, which is an anomaly detection algorithm that can provide an alert when malicious activity, such as data leakage via a covert channel, takes place over the bus.

A related domain to our research is the inter-communication technologies implemented in today's automobiles, also known as the CAN bus. Numerous works surveyed cars' inter-communication technologies and possible threats \cite{wolf2004security,klegerger2011security}, and demonstrated different kinds of attacks \cite{hoppe2008security,checkoway2011comprehensive,roufa2010security,miller2015remote}.
These works suggested adding countermeasures such as encryption, authentication, and intrusion detection capabilities to these technologies.
Some examples of proposed anomaly detection solutions for identifying malicious activity on CAN bus, include: 
an entropy-based anomaly detection method that assumes low randomness in the vehicle network  \cite{muter2011entropy}; an unauthorized data transmission prevention mechanism for CAN bus \cite{matsumoto2012preventing}; denial of service detection by analyzing time intervals of messages \cite{song2016intrusion}; an intrusion detection solution that is based on clock skew fingerprints of computational unites \cite{cho2016fingerprinting}; and an intrusion detection solution based on a supervised DNN classifier used for classifying messages as benign or anomalous by analyzing their content \cite{kang2016DNN}.

Although the CAN bus protocol and physical implementation differ from the MIL-STD-1553 communication bus, some of the suggested methods and features for anomaly detection can be relevant for MIL-STD-1553, as will be described in Section \ref{sec:seq-timing-module}.

\section{\label{sec:Security-Analysis-of}Security Analysis of the 1553 Communication Protocol}
In this section we present a comprehensive security analysis of the MIL-STD-1553 communication protocol which consists of the following elements:

\textbf{Assets} -- an element, which is part of the 1553 communication bus, that (1) an attacker might be interested in, and (2) has the potential to disrupt the system's operation or leak information when compromised. An asset might be a physical component (e.g., a subsystem), or data present in the system (e.g., transmitted messages, data stored in a subsystem).

\textbf{Attacker~profile} -- an individual, group, organization, or government that have interest in attacking the system's assets and attempt to access them via attack vectors.
	
\textbf{Attack~Vector} -- indicating various methods used by an attacker to penetrate the system in order to perform the malicious activity.
	
\textbf{Threat~Agent} -- an entity (individual, software, hardware), internal or external to the system, that uses its privileges in order to execute the attack.

\textbf{Attack~method} -- the actions that an attacker should perform in order to execute an attack.

\subsection{Assets}

The identified assets that are part of a MIL-STD-1553 communication bus and might have value to a potential attacker are can be categorized as follows:

\begin{description}
	\item[Connectivity assets] -- the physical components responsible for data transfer between the different components at different levels: 
		\begin{itemize}[noitemsep,nolistsep]				
			\item Transmission medium (the bus itself) (component 8 in Figure \ref{fig:MIL-STD-1553-bus-architecture}) -- the physical wires that connect the RTs and enable all communication and data transmission.
			
			\item Transceiver (component 2 in Figure \ref{fig:MIL-STD-1553-bus-architecture})
			-- responsible for decoding the analog signals into digital data which is comprehensible to the subsystem (and vice versa) and thus enables data transfer between the bus and the subsystem.
			
			\item Coupler (component 7 in Figure \ref{fig:MIL-STD-1553-bus-architecture}) 
			-- an electrical unit that isolates the bus from an RT and connects the transceivers to the transmission medium.
		\end{itemize}
	Damage to one of these components might harm the availability of a part, or the entire, system. Denying a critical subsystem to transmit data (by sabotaging its connectivity assets for example) prevents inputs for other component that might fail in performing their tasks, potentially leading to disconnection between components. Since these components are physical, they are capable for compromising emanation, which harms the confidentiality of the system. The integrity of the system is also threatened by these assets, since they have access to the inputs and outputs of each component, and once compromised, they can manipulate these data.
	
	\item[Data assets] -- the data stored in different parts of the system:
		\begin{itemize}[noitemsep,nolistsep]
			\item DPR data (component 3 in Figure \ref{fig:MIL-STD-1553-bus-architecture})
			-- the data stored in the shared memory of the transceiver and the subsystem (DPR).
		 
		 	\item Subsystem data (component 4 in Figure \ref{fig:MIL-STD-1553-bus-architecture})
			-- the data that is stored in the memory of the subsystem and consumed by the subsystem in order to perform its tasks (e.g., geographical location).
		
			\item Data in motion (components 8 in Figure
			\ref{fig:MIL-STD-1553-bus-architecture}) -- the current signals (data) transmitted over the bus.
		\end{itemize}
	Any damage or changes made to these assets violates the integrity of the system. Moreover, as previously described, manipulation of the inputs and outputs of subsystems can damage the system's availability. Moreover, lack of data encryption breaks the system's confidentiality once leaked outside.
	
	\item[Computational units:]~\newline \vspace{-\baselineskip}
		\begin{itemize}
			\item Subsystem (component 4 in Figure \ref{fig:MIL-STD-1553-bus-architecture})
			-- consists of physical components (e.g., CPU, memory, sensors) and the software responsible for performing the subsystem's tasks (e.g., reporting the current position, calculating distance from objects).
		\end{itemize}
	Compromised subsystems can manipulate or generate false outputs and break the system's integrity, stop communication with other subsystems and damage its availability, or abuse access to other devices in order to leak data and violate the system's confidentiality.
\end{description}

Table \ref{tab:list-of-assets} provides more detailed description of each asset and the security concerns (i.e., potential consequences) related to it, categorized by integrity, confidentiality, and availability.

\subsection{Attacker profile}

Since MIL-STD-1553 is mainly implemented in military platforms, most of the attack vectors require physical access to the system (e.g., change components' code, eavesdropping), or access to  external devices that interact with the system (such as USB devices or CDs) or sensors (such as GPS or RADAR). This kind of access requires highly skilled attacker such as a state actor.

The attack vectors can be categorized into three main groups: code injection and manipulation, data injection, and physical tampering. The attacker is assumed to have the ability to execute at least one of these attack vectors during the life cycle of the system (e.g., development, supply chain, deployment, or maintenance stages). These individuals can abuse their access rights in order to sabotage various components.

Once the attacker gained access to the system he/she executes the attack via a component connected to the system or an individual that have physical access to it, which are referred to as \emph{threat agents}.

\subsubsection{\label{subsec:attack-vectors}Attack vectors}
\textbf{Code~injection~and~manipulation.} This attack vector refers to the ability to inject or manipulate the code of the system's components
in order to perform the attack. This includes the program coded in the transceiver, as well as the operating system or software of a subsystem. Malicious code can be injected during the main phases of the component's life cycle: development, supply chain, and deployment and maintenance.

The development phase includes all processes that take place before delivering the 
product to the client: hardware
\begin{table}[h]
	\begin{minipage}{\textwidth}
		\centering
		\caption{List of assets and potential consequences.}
		\label{tab:list-of-assets}
	\end{minipage}
		\renewcommand{\arraystretch}{1.5}
			\begin{tabular}{|>{\centering}m{0.005\textwidth}|>{\centering}m{0.15\textwidth}|>{\centering}m{0.25\textwidth}|>{\centering}m{0.25\textwidth}|>{\centering}m{0.25\textwidth}|}
				\hline
				& {\textbf{Asset}} & {\begin{center}\textbf{Integrity}\end{center}} & {\begin{center}\textbf{Confidentiality}\end{center}} & {\begin{center}\textbf{Availability}\end{center}}\tabularnewline
				\hline
				1 & Transceiver  & \raggedright Compromised transceiver can provide corrupted data to the subsystem it connects to the bus or to other components connected to the bus which can lead to incorrect operation. &  & \raggedright Compromised/corrupted transceiver can stop data transfer between the bus and the subsystem which can lead to DoS to the subsystem it connects to the bus, and/or to other components that depend on the data it should transmit.
				\tabularnewline
				\hline 
				2 & Transmission medium (the bus itself)  & \raggedright Shorts or failure of the transmission medium may provide corrupted data to the components connected to the bus which can further lead to incorrect operation
				of the system. & \raggedright Electromagnetic energy emanating from compromised transmission medium may be used to deduce the information transmitted on the bus and compromise the system's confidentiality. & 
				\raggedright Shorts or failure of the transmission medium may lead to total disconnection of the communication over the bus and interrupt the system's operation. 
				\tabularnewline
				\hline 
				3 & Coupler & \raggedright Compromised coupler can provide corrupted data to the RT it connects to the bus, or to other components connected to the bus which can lead to incorrect operation. & 
				\raggedright Electromagnetic energy emanating from a compromised coupler may be
				used to deduce the information transmitted on the bus and compromise the system's confidentiality. & 
				\raggedright Unavailable coupler disconnects the RTs connected to the coupler from
				the bus. In some cases it can also cause DoS to other components connected to the bus. 
				\tabularnewline
				\hline 
				4 & Subsystem & \raggedright Compromised subsystem can provide corrupted data to other components and lead to incorrect operation. It can also spoof as another component by changing the TA field of a command.  & \raggedright Compromised subsystem can abuse access to devices that have the ability
				to transmit data outside the system (i.e., radio transmitter) and leak
				sensitive information.  & 
				\raggedright Unavailable subsystem stops responding to commands and data transmission which might lead to DoS to other components depending on its outputs and possibly even to the entire system. Corrupted data sent by a compromised subsystem to other subsystems may also result in DoS.
				\tabularnewline
				\hline 
				5 & DPR data & \raggedright Corrupted data provided to a component can lead to incorrect operation. & 
				\raggedright Classified data that leaks outside the system in plain text can be abused by malicious individuals. & 
				\raggedright Unavailable or corrupted data may lead to DoS to the components depending on it, and possibly even to the entire system.
				\tabularnewline
				\hline 
				6 & Subsystem data & \raggedright Corrupted data provided to a component can lead to incorrect operation of the system. & 
				\raggedright Classified data and/or operation logic that leaks outside the system in plain text can be abused by malicious individuals. & 
				\raggedright Unavailable or corrupted data may lead to failure of the subsystem's
				operation and may also result in DoS to the components depending
				on its outputs and possibly even to the entire system.\tabularnewline
				\hline 
				7 & Data in motion & \raggedright Corrupted data provided to a component can lead to incorrect operation of the system. & 
				\raggedright Classified data and/or operation logic can be leaked outside the system by compromising emanation and can be abused by malicious individuals. & \raggedright Unavailable data might lead to DoS to the components depending on
				it and possibly even to the entire system.
				\tabularnewline
				\hline 
		\end{tabular}
\end{table}
	\noindent% here to save space
manufacturing, code writing, integration, and testing. 
During the development phase, malicious individuals can exploit their access to the components and insert erroneous or malicious code, or physically tampering with components. Though components are tested before they are delivered to clients, a sophisticated attacker can inject code that is programmed to operate
within a specific context and can identify when it is in the real environment, thereby evading detection tools.

During the deployment and maintenance phase various procedures performed may expose the system to malicious code injection. 
These procedures include: operating system and software updates, bug fixes, system configuration, and data loading. Such maintenance activities may be performed via wireless communication or physically via CD/DVD, USB connection, or through a computer that is connected to the bus. In this phase, code injection and manipulation may also be performed by another component that was previously compromised and is connected to the bus.

\textbf{False~data~injection.} Data injection refers to false data provided by sensors, such as Global Positioning System (GPS) or Radio Detection And Ranging (RADAR) systems, or an external device (e.g., magnetic tape, CD/DVD, or computer). In recent years extensive research has been conducted regarding false data injection attacks on control systems, mainly on electrical power grids \cite{liu2011false,mo2010false}. 
In this type of attack, the attacker injects crafted data into the system through sensors (or other input devices) that alter the normal behavior of the system and might lead to failures and even the execution of malicious code. Note, however, that in order to perform a successful data injection attack without detection, the attacker must have in depth knowledge of the system and its vulnerabilities.

\textbf{Physical~tampering.} Every electronic device emits electromagnetic radiation. By eavesdropping on the device and analyze its electromagnetic emanations, an attacker can reveal information regarding the device's operation. This type of attack is called tempest \cite{kuhn1998soft}, and it is addressed in the Designer's Notes for MIL-STD-1553.
However, a malicious individual who has physical access to the system can make subtle changes to the system, such as adding computational capabilities to a coupler, manipulating wiring or the coupler's grounding. Such modifications may not change the component's behavior significantly, but does create some type of side effect (such as amplified electromagnetic radiation), which may also go undetected
if the system is not specifically tested for those specific side effects. 

\subsubsection{Threat agents}
After the attacker managed to gain access to the system, he/she can use one of the following threat agents in order to execute an attack: a component connected to the bus or a malicious individual (human) possessing access permissions to the system.

\textbf{Component~(RT~and~BC)} -- we distinguish between two types: a compromised component and a fake component. 
A compromised component is a component which was originally part of the system and was manipulated by the attacker. This may include components that are not constantly connected to the bus and are connected on demand (e.g., for uploading configurations, downloading logs, and maintenance). 
A fake component was not part of the system and was connected to it illegitimately. Once connected, the fake component becomes part of the system and can transmit data and listen to all communications. We also distinguish between BC and RT components, since the functionality of the BC is more extensive than the functionality of the RT, and hence has greater capabilities for executing attacks.

\textbf{Malicious~individual} -- an individual (human) that cooperates with the attacker (or the attacker himself) and has access to the system, who can tamper with its components physically (by sabotaging their circuitry, for instance), or logically (e.g., by inserting errors in a component's code).

\subsection{\label{subsec:threats}Attack Methods and Consequences}

This section describes the threats to the MIL-STD-1553 communication protocol, which are categorized by their impact: denial of service, data leakage and data integrity violation. The following subsections elaborate on the different threats to the 1553 communication bus and provides methods to execute them. Table \ref{tab:threats-methods} provides more detailed description of each attack method, categorized by message manipulation and behavior manipulation. Message manipulation refers to modification of legitimate words (command, data, or status) transmitted over the bus. Behavior manipulation refers to altering the behavior of the compromised component, for example, transmitting fake (malicious) messages in unusual timings or order.

\subsubsection{Denial of Service (DoS)} DoS can be achieved by damaging physically or logically the system's assets, and will usually require only one threat agent. Physical damage to a component can harm its ability to perform operations, produce outputs, or transmit them over the bus. In particular, if the damaged component is the bus itself, there could be a complete disconnection between all of the components connected to it.

Logical damage refers to exceptions occurred during component's normal operation, component's incorrect operation, or data manipulation and corruption. These scenarios result in corrupted output or lack of response, which can lead to denying the operation of one or more components. Following a description of possible methods to achieve DoS to an 1553 communication bus.

\textbf{Message manipulation:} compromised components with BC capabilities can change fields of a command word (e.g. WC, T/R, and TA) to control data routing and cause collisions. For example: denying a 'transmit' command from reaching the GPS by changing its TA field will cause other subsystems (e.g. navigation, artillery, etc.) rely on outdated data, which can have severe outcomes. Manipulating status words by a threat agent to falsely indicate on errors in the target RT might lead to termination of the communication with it, although it operates correctly. Data words can be easily corrupted by different threat agents, by causing collisions or manipulating them at the subsystem's level. Lack of sufficient input validation by the component can lead to an incorrect operation and even crash it.

\textbf{Behavior manipulation:} compromised components that can control their transmission times and response delays or behave differently than the command specifies can also cause collisions and failures to other component, thus lead to DoS.

\textbf{Possible operational consequences:} DoS to the 1553 communication bus can have devastating results, especially because it is used for mission critical systems. The attack can be executed upon detection of some operation in order to intercept it. For example, an attacker that listens to the bus can identify that the system entered a certain geographic zone and deny location data from updating the navigation system, or identify that the system is aiming to fire at a target and deny the firing command from reaching the relevant components.

\subsubsection{Data leakage} Data leakage in the context of the MIL-STD-1553 communication bus is the result of unauthorized data transmission between components (i.e. components of different security levels) or outside the system.

\textbf{Message manipulation:} by changing the WC or TA fields in a command word, a threat agent can instruct a component to transmit exceeding data words, or to transmit data words to another component (that may have lower security level). Data can also be leaked using the reserved bits of a status word, or by modulating additional payload on legitimate data words.  

\textbf{Behavior manipulation:} threat agents that can control their behaviors are capable of creating a covert channels in order to leak data as presented by Neugen in \cite{nguyen2015towards}. If the threat agent has BC capabilities it can also utilize idle time on the bus and initiate unauthorized data transfers. Moreover, if the threat agent has an access to an external device or removable hardware it can utilize it to leak data outside the system. Data can also be leaked physically by eavesdropping the electromagnetic emanations of components.

\textbf{Possible operational consequences:} Leaked data can help the attacker conclude information about the operation of the system. Usually vehicle have service ports (e.g. USB) that are easily accessible to maintenance crew for debugging and investigating the vehicle's performance. A malicious crew member can extract logs and traffic traces from the system and pass them to the attacker. Sensitive information, such as: current vehicle location, targets, and destinations, can be leaked by a compromised component outsize the vehicle by using legitimate external communication channels (e.g., radio).

\subsubsection{Violation of data integrity} Violation of data integrity refers to invalid or incorrect data that flows inside the system and causes other component to fail or operate incorrectly. Incorrect data can get inside the system by a threat agent external to the system (see \ref{subsec:attack-vectors}) or by an inside threat agent that can manipulate messages exchanged over the bus or send fake data in the behalf of another component (i.e., spoofing), and cause the system to behave abnormally.

\textbf{Possible operational consequences:} by altering the data words an attacker can cause the system to navigate to the wrong destination, fire at the wrong target, and even to no fire at all, or withhold/add objects from/to the vehicle's dashboards and deceive the crew aboard.

\section{\label{sec:seq-timing-module}Sequence-Based Anomaly Detection Method for the 1553 Communication Bus}

To the best of our knowledge, there are no security solutions for identifying and/or preventing cyber attacks on the 1553 communication bus. Existing security solutions (e.g., firewall, intrusion and malware detection, data leakage prevention, access control) are not suitable for the 1553 communication bus because they require significant adaptation and configuration to the specific operating systems and communication protocol. 
Existing solutions also require the application of changes to various components of the 1553 communication bus which may be cost prohibitive due to its extensive deployment in various aircrafts and vehicles.

Therefore, we propose adding a lightweight, MIL-STD-1553 tailor-made anomaly detection \cite{chandola2009anomaly} solution that is based on continuous monitoring of the messages transmitted over the bus and the application of machine learning-based anomaly detection algorithms, in order to identify attacks on the bus. In this section we present a sequence-based anomaly detection module, which is a solution for identifying command word manipulations and timing-related behavior manipulations.

The advantages of the proposed solution are two-fold. First, by using machine learning techniques, which are highly flexible and adaptive \cite{garcia2009anomaly}, we can provide a robust solution which can be automatically adapted to any 1553-based system in a very short time. Second, the proposed solution can be implemented (integrated) as part of a Bus Monitoring (BM) module and therefore does not require any change to the existing modules of the bus.

The sequence-based anomaly detection module focuses on detecting whether the message complies with the predefined major frame specification for the specific bus implementation, or whether it arrived out of order or was sent at the wrong time.
Similar to \cite{cho2016fingerprinting}, we use time interval analysis of messages and inspect their deviation from their normal time cycle.

Since most of the messages sent over the bus are periodic, it is more likely that command and timing features will be useful for identifying anomalous messages. For that, we propose the use of sequence mining algorithms, such as a Markov chain model, which are able to derive a model which represents valid transitions of messages from a training set containing legitimate messages.

\begin{figure}[h]
	\includegraphics[width=0.52\textwidth,center]{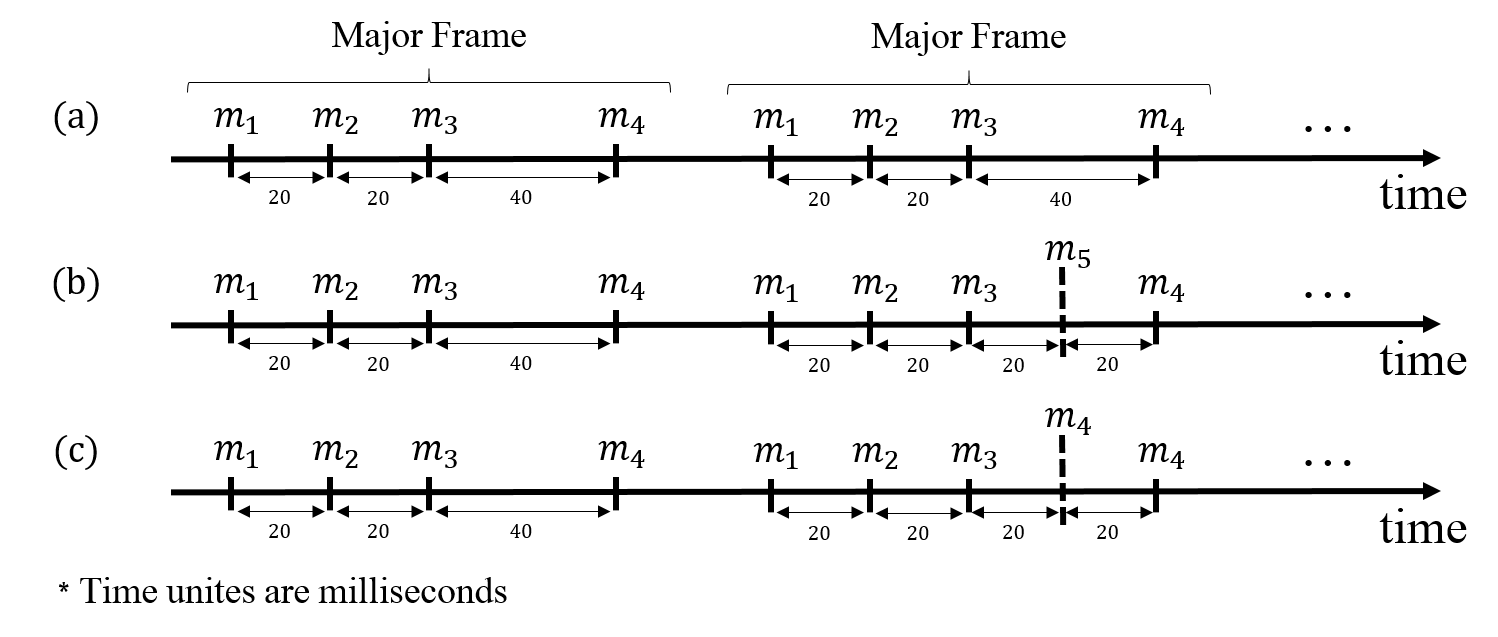}
	\par
	\vspace{-16pt}
	\caption{\label{fig:SequenceModelExampleScenario}Simple major frame examples for: (a) a legitimate major frame, (b) new message injection attack, (c)  legitimate message injection attack.
	}
\end{figure}

Models which are based on command features can help detect messages that are sent out of order. For example, consider a simple case of a major frame that consists of four messages: $m_{1},m_{2},m_{3}$ and $m_{4}$ that are sent as illustrated in Figure\ref{fig:SequenceModelExampleScenario}(a). If an attacker tries to utilize the idle time between ,$m_{3}$ and $m_{4}$ to send its own message - $m_{5}$ (Figure \ref{fig:SequenceModelExampleScenario}(b)), the model will identify it immediately as an anomaly, since the only acceptable transition from $m_{3}$ is to $m_{4}$. However, if the attacker chooses to utilize $m_{4}$ (Figure  \ref{fig:SequenceModelExampleScenario}(c)), it might be detected as benign and cause the real $m_{4}$ to be detected as an anomaly. In this case the timing features will assist the model in detecting the anomalous command, since the attacker's $m_{4}$ will be sent at a timing that does not comply 
\begin{table}[!t]
		\begin{minipage}{\textwidth}
			\centering
			\caption{Threats and attack methods}
			\label{tab:threats-methods}
		\end{minipage}
	\fontsize{6.3}{6.3}\selectfont 
		\renewcommand{\arraystretch}{1.5}
			\begin{tabular}{|c|>{\centering}m{0.1\textwidth}|>{\raggedright}m{0.27\textwidth}|>{\raggedright}m{0.26\textwidth}|>{\raggedright}m{0.26\textwidth}|}
				\hline
				& \textbf{Category} & \centering \textbf{DoS attack} & \centering \textbf{Data Leakage} & \centering \textbf{Data integrity vaiolation}\tabularnewline
				\hline 
				\multicolumn{5}{|c|}{\textbf{Message manipulation}} \tabularnewline
				\hline
				1 & Command word  & 
				\textbf{WC field}
				\begin{itemize}[noitemsep,nolistsep,wide=0pt]
					\item[--] Changing the WC field to a smaller number
					causes the target RT to receive or transmit partial data which can lead to an error in the target RT or other RTs depending on its output.
					
					\item[--] Changing the WC field to a larger number can also lead to an error due to collisions and corrupted data reception.
				\end{itemize}
				\textbf{T/R bit}
				\begin{itemize}[noitemsep,nolistsep,wide=0pt]
					\item[--] Flipping the T/R bit in a 'transmit' command causes the target RT to receive a 'receive' command that causes the target RT to respond with an error or wait for data to arrive (while no data is transmitted); furthermore the RT won't send the data it should to other RTs and they won't get their inputs.
					
					\item[--] Flipping the T/R bit in a 'receive' command causes the target RT to receive a 'transmit'  command that can lead to an error or data transmission that causes collision (since the BC continues to transmit the data of the 'receive' command).
				\end{itemize}
				\textbf{TA field}
				\begin{itemize}[noitemsep,nolistsep,wide=0pt]
					\item[--] Changing the TA field to another/unsupported RT address prevents the command from reaching its target RT and can cause a failure in the RT's operation or failure of other RTs depending on it.
				\vspace{-\baselineskip}
				\end{itemize}  &
				\vspace{-47pt}
					
				\textbf{WC field}
				\begin{itemize}[noitemsep,nolistsep,wide=0pt]
					\item[--] By changing the WC field of a 'transmit' command to a larger number the threat agent might cause the target RT to transmit more data than it should. If the attacker is familiar with the memory map of the target RT, he/she can use this method to access restricted areas in the target RT's memory.
				\end{itemize}
				\textbf{TA field}
				\begin{itemize}[noitemsep,nolistsep,wide=0pt]
					\item[--] By changing the TA field in a 'transmit' command to another RT address the threat agent might obtain data from a subsystem that it is not authorized to hold.
					
					\item[--] By changing the TA field of a 'receive' command, the threat agent can force an RT to accept data that it might not be authorized to hold.
			
			\end{itemize}  &
				\vspace{-84pt}
				A threat agent with BC capabilities can be used to tamper with the communication between the real BC and various RTs. The threat agent can corrupt the original command when it is transmitted over the bus, and send its own command to the target component instead. The target component will send its response without knowing that the command received is different than the original one, and the real BC will receive a response for a command it did not send.
				
				\tabularnewline
				\hline
				2 & Status Word  & 
				
				A compromised RT can impersonate as another and set the 'Busy', 'Terminal', or 'Subsystem' flags in its status word and provide a falsely indication to the BC regarding a malfunction or inability to handle messages and thus disrupt the communication with that RT. Similarly, a fake BC can respond on behalf of the target RT and signal the BC to stop sending commands to the target component.  & 
				
				Leaking data via status words can be done by utilizing the 'reserved' bits (see Figure \ref{fig:Communication-protocol-words}) - three bits that are reserved for future development of the standard.
				The standard specifies that these bits should be unused and remain set to zero. A lack of status word monitoring enables cooperating threat agents to easily transfer any data without detection.  	& 
				
				\vspace{-18pt}
				Any threat agent connected to the bus (with BC or RT capabilities) can corrupt status words transmitted back to the real BC and send fake statuses as if is the transmitting RT.
				
				\tabularnewline
				\hline
				3 & Data Word  & 
				
				\vspace{2pt}
				\begin{itemize}[noitemsep,nolistsep,wide=0pt]
				\item[--] A malicious BC or RT can alter legitimate data transmitted and cause failure in the target component (if the target component doesn't perform validation at the subsystem level).
				\item[--] An attacker who has prior knowledge about the target component can generate and inject fake data that can cause failure, disrupt the normal operation, or impair the outputs of the target component.
				\vspace{-\baselineskip}	
				\end{itemize}
				& 
				
				\vspace{-18pt}
				Any threat agent can use the data words it transmits in order to modulate additional payload. This type of attack requires a cooperating threat agent who is familiar with the modulation method and can then decode the additional payload.
				& 
				Threat agents can utilize idle times on the bus and resend fake commands to target components on behalf of legitimate components, in order to override the real data stored in the target components' memory. The target components will consider the fake data to be the real data received from the legitimate component.
				
				\tabularnewline
				\hline
				\multicolumn{5}{|c|}{\textbf{Behavior manipulation}} \tabularnewline
				\hline
				
				4 & Command Word  & 
				
				\textbf{Fake command}\newline
				Issuing fake commands (either defined by the standard or meaningless) that are not part of the system's normal operation  may result in collisions, blocking all communication over the bus or affecting the proper system's operation (e.g. issuing shut-down commands or clock synchronizing at incorrect timings).
				
				\textbf{WC field}
				\begin{itemize}[noitemsep,nolistsep,wide=0pt]
					\item[--] Sending less data than specified by the WC field of a command causes the target component to receive incomplete data and may fail to operate. 
					
					\item[--] Sending excessive amount of data can cause a collision if the target component responds with its status while the threat agent is still transmitting data.
				\vspace{-\baselineskip}	
				\end{itemize}
				&  
				
				\vspace{-55pt}
				Neugen presented in \cite{nguyen2015towards} a storage attack method to create covert channel between two compromised components of different security levels over the 1553 bus, which requires a compromised BC and a compromised RT, and is based on the RT's specific 'command illegalization' implementation.
				& 
				
				\tabularnewline
				\hline
				
				5 & Status word  & 
				
				&  
				
				Neugen presented in \cite{nguyen2015towards} a storage attack method to create covert channel between two compromised components of different security levels over the 1553 bus, which is based on the Service Request (SR) bit of a status word and requires a cooperating BC and RT.
				& 
				
				\tabularnewline
				\hline
				6 & Transmission timings  &
				
				\vspace{-27pt}
				\begin{itemize}[noitemsep,nolistsep,wide=0pt]
					\item[--] Threat agents that can control the timing of their transmissions can transmit messages at the time of choice. Sending unexpected messages to target components may result in failures.
					
					\item[--] Threat agents that can control the timing of their transmissions can cause collisions that corrupt data transmitted over the bus (e.g., by transmitting at random timing) and can lead to error or incorrect operation of other components.
				\vspace{-\baselineskip}	
				\end{itemize}
				
				&  
				\vspace{2pt}
				\begin{itemize}[noitemsep,nolistsep,wide=0pt]
					\item[--] Neugen presented in \cite{nguyen2015towards} a timing attack method to create covert channel between two compromised components of different security levels over the 1553 bus,  
					in which two cooperating RTs establish a signaling mechanism based on their response time delays that are interpreted into binary data.
					
					\item[--] Threat agent with BC capabilities can utilize idle time periods on the bus and initiate data transfer with any RT in order to extract data. If there is a cooperating threat agent connected to the bus, then the agent with BC capabilities can initiate RT-RT communication and transfer data from the target RT to the cooperating threat agent.
				\vspace{-\baselineskip}	
				\end{itemize}
				& 
				
				\tabularnewline
				\hline
				
				7 & BM impersonation   &
				&  
				
				Any threat agent connected to the bus can act as a BM and record the data transmitted over the bus which
				is available to all components connected to the bus. This data may be further leaked to other components or external devices via removable hardware (e.g., USB, CD, or magnetic tape), an available connection to other networks, or covert channels.
				&
				
				\tabularnewline
				\hline
				8 & tempest    & 
				
				&  
				Malicious individuals can eavesdrop and capture the electromagnetic emanations of components \cite{kuhn1998soft} (which can be enhanced by physically sabotaging the components), and analyze them in order to obtain information about the target component's operation that can imply on other operations and characteristics of the entire system and help the attacker better understand it. & 
				
				\tabularnewline
				\hline 
		\end{tabular}
\end{table}\noindent	% here to save space
with its normal transmission timing, as defined by the major frame.

If a message is found anomalous by this module, an alert is generated.

\vspace{-8pt}
\subsection{\label{subsec:Detection-algorithm}Detection algorithm}

As previously described, we detect anomalous messages based their timing and order. We opt to use the Markov chain model as the basis for our
detection mechanism, in order to represent the normal behavior of 
the monitored bus (similar to \cite{ye2000markov}). 

We distinguish between two types of messages: periodic and aperiodic. A periodic message is sent by the BC
at fixed time intervals (referred to as time cycles). An aperiodic message is sent by the BC
 as a result of an event or a Service Request. 

In order to profile both periodic and aperiodic messages we use two Markov models - one for each type. 
The \emph{periodic} Markov model's states are defined by command and
timing features (see Table \ref{tab:List-of-extracted}).
Aperiodic messages, which are sent upon demand, are likely to be found anomalous by the periodic model which uses time-related features. 
Therefore, a second model is maintained, the \emph{aperiodic} Markov model, in which each of the states are represented only by command features. Next, we provide a detailed description of the training and detection processes of the proposed method.

\begin{table}[h]
	\caption{Extracted features used for defining the Markov model states}
	\label{tab:List-of-extracted}
	\fontsize{6}{7}\selectfont
	\renewcommand{\arraystretch}{1.2}
	\centering
		\begin{tabular}{|>{\centering}m{0.05\textwidth}|>{\centering}m{0.05\textwidth}|>{\centering}m{0.05\textwidth}|>{\raggedright}m{0.22\textwidth}|}
			\hline 
			& Feature name & Values & \centering Description\tabularnewline
			\hline 
			\hline 
			\multirow{7}{0.05\textwidth}{Command features} & Src. Terminal Address  & 0-31, N/A & The address of the terminal sending the data.
			If the terminal is the BC, the address is 'N/A'.\tabularnewline
			\cline{2-4} 
			& Src. Subaddress & 0-31, N/A & The subaddress from which the data is sent in the source terminal. If the terminal is the BC, the subaddress is 'N/A'.
			\tabularnewline
			\cline{2-4} 
			& Dst. Terminal Address & 0-31, N/A & The address of the terminal receiving the data. If the terminal is the BC, the address is 'N/A'.\tabularnewline
			\cline{2-4} 
			& Dst. Subaddress & 0-31, N/A & The subaddress to which the data is saved in the destination terminal. If the terminal is the BC, the subaddress is 'N/A'.\tabularnewline
			\cline{2-4} 
			& Channel & A, B & The channel on which the message was sent.\tabularnewline
			\cline{2-4} 
			& Word Count & 0-32 & The number of data words sent in the message frame.\tabularnewline
			\cline{2-4} 
			& Is Mode Code & true, false & Whether the command is a mode code or not.\tabularnewline
			\hline 
			{Timing features} & Time Cycle & numeric & The time cycle (in microseconds) of the message. Note that a message can have several different time cycles.
			\tabularnewline
			\cline{2-4}
			\hline 
			& Label & Benign, Anomaly, N/A  & The message label: 'Benign', 'Anomaly', or 'N/A' (for cases the model cannot classify the message).\tabularnewline
			\hline 
		\end{tabular}
\end{table}

\subsubsection{Training process}

The training phase requires a training set that consists solely of normal bus operation (i.e., legitimate messages). During this phase, the states of each of the two Markov models are identified, and then the states' transition probabilities are computed.

\textbf{Identifying~the~Markov~model~states.} Each $state_{j}$ in the Markov model is defined by a set of features that are extracted from the messages observed in the training set (denoted by $TS$). 
The states of the aperiodic Markov model are defined by the following seven features (listed in Table \ref{tab:List-of-extracted}): 'Src. Terminal Address', 'Src. Subaddress', 'Dst. Terminal Address', 'Dst. Subaddress', Channel, 'Word Count', and 'Is Mode Code'.

We refer to each unique tuple (defined by the values of the above features) as a message ID. Each message ID defines a specific state of the aperiodic Markov model.
In addition to the above seven features, the states of the periodic model include an eighth feature -- the 'Time Cycle' feature -- which indicates the time cycle of a periodic message. Since a periodic message can appear in one or more predefined time cycles, there is a need to deduce the time cycle(s) of the message in order to define the states of the periodic Markov model.

Therefore, in order to classify a message as periodic or aperiodic, and to define the states of the periodic Markov model, for each message ID we compute the 'Time Cycle' feature as described in Algorithm \ref{alg:identfiy-time-cycles}. Let $TS_{j}$ be the subsequence of messages containing the instances of message ID $j$. For each message in $TS_{j}$ we calculate the sequence of time differences between two consecutive appearances of the message (referred to as 'time difference sequence' and denoted by $TD_{j}$ in Algorithm \ref{alg:identfiy-time-cycles}).
The time difference sequence is ordered in ascending order,
and is then clustered in a greedy manner as follows.
Each time difference value in the ordered sequence is compared with the previous value. If the difference between the current value and the previous one is smaller than a predefined threshold (denoted by $tr$), the current time difference is assigned to the current cluster; otherwise, a new cluster is created with the current time difference value. 
In our study we set the predefined threshold to be 40 microseconds which is the acceptable deviation in the time difference values in existing platforms implementing the 1553 communication bus.
Eventually each cluster represents a time cycle of message ID $j$ and the representative value of each cluster is the average of the time difference values in the cluster.

After extracting the time cycles of each message ID we can classify them as periodic and aperiodic by analyzing their time cycles and number of occurrences as follows.
If a message ID is very rare (i.e., the number of times it was observed is lower than a predefined threshold), it is classified as aperiodic.
If the number for detected time cycles (according to Algorithm \ref{alg:identfiy-time-cycles}) of a message ID is greater than a predefined threshold (set to three in our study) it is classified as aperiodic. In any other case the message ID is classified as periodic.

\begin{algorithm}[h!]
	\caption{Extract time cycles of a message}
	\label{alg:identfiy-time-cycles}
	\begin{algorithmic}[1]
		\Procedure{ExtractTimeCycles}{$TS_{j}$}
		\For{$i:=1~to~|TS_{j}|$}
		\State $TD_{j}.append(TS_{j}[i].time - TS_{j}[i-1].time)$ 
		\EndFor
		\State $Sort(TD_{j})$
		\State $k\gets 0$
		\State $clusters[k]\gets TD_{j}[k]$
		\For{$i:=1~to~|TD_{j}|$}
		\State $diff\gets TD_{j}[i] - TD_{j}[i-1]$
		\If{$diff > tr$}
		\State $k\gets k + 1$
		\EndIf
		\State $clusters[k].append(diff)$
		\EndFor
		\For{$i:=0~to~k$}
		\State $cycles_{j}[i]\gets ComputeAverage(clusters[i])$
		\EndFor
		\Return $cycles_{j}$
		\EndProcedure
	\end{algorithmic}
\end{algorithm}

\textbf{Computing~the~state~transition~probabilities.} The second step in the training phase, after identifying the states of the periodic and aperiodic Markov models, is computing the state transition probabilities. By iterating over the training set we first compute the following values for each model: 

\begin{itemize}[noitemsep,nolistsep]
	\item {$occur_{j}~=~count(state_{j})$ -- } the number of times that $state_{j}$ was observed in $TS$.
	\item {$trans_{j\rightarrow l}~=~count(state_{j}\rightarrow state_{l})$ -- } the number of times in which  $state_{l}$ appeared after $state_{j}$ in $TS$.
\end{itemize}

Then we can calculate the following probabilities:
\begin{itemize}[noitemsep,nolistsep]
	\item $stateProb_{j} = \frac{occur_{j}}{|TS|}$ -- probability of observing $state_{j}$ in $TS$.
	\item $transProb_{j\rightarrow l}=\frac{trans_{j\rightarrow l}}{\sum_{k\in S}{}trans_{j\rightarrow k}}$
	-- probability of transition from $state_{j}$ to $state_{l}$ ($S$ is the set of all states discovered in $TS$).
\end{itemize}

Note that when calculating probabilities for the periodic model, transitions involving aperiodic messages are ignored (i.e., not considered in the computation of probabilities). This is because aperiodic messages cannot be mapped to the states of the periodic model (i.e, they do not have a time cycle).

\textbf{Calculating~the~anomaly~threshold.} During the detection phase, the derived Markov models assign a probability for an observed transition of two messages. In order to classify the observed transition as normal
or abnormal, the probability is compared against an anomaly threshold.
The anomaly threshold (denoted by $tr_{a}$) is defined as the minimal probability of a sequence of length two observed in the training set.
The probability of the sequence $[m_{1}$,$m_{2}]$ is computed as follows: $stateProb_{s_{1}}\cdot transProb_{s_{1}\rightarrow s_{2}}$, where $s_{i}, (i=1,2)$ is the corresponding state of $m_{i}$ in the relevant model.

\subsubsection{\label{subsec:detection-process}Detection process}
In the detection phase, a message is examined in order to see if it was manipulated in a specific way. The input to the detection algorithm (Algorithm \ref{alg:Detecting-order-timing-anomalies}) is the inspected message ($msg_{i}$) and the relevant model ($model$), which is determined according to the message's type (periodic or aperiodic).

First, the model is applied on $msg_{i}$ (by evaluating the transition from the previous message $msg_{i-1}$ to the current message $msg_{i}$) in order to compute the anomaly score (line 2, as in the \emph{anomaly threshold} step).
If the anomaly score is greater than or equal to the model's threshold ($tr_a$), $msg_{i}$ is labeled as 'Benign' (lines 3-4). Otherwise $msg_{i}$ is labeled as 'Anomaly'.

A message that was classified as 'Anomaly' (e.g., a crafted message that was injected by the attacker) can lead to a misclassification of its successive benign messages. This is because the transition from an anomalous message $m_{a}$ to its successive message $m_{b}$ will be detected as anomalous even if $m_{b}$ is benign since the transition $s_{a}\rightarrow s_{b}$ (the corresponding states) is not recognized by the model.
Therefore, in order to avoid incorrect classification of successive messages we apply (in lines 7-10) \emph{a point anomaly recovery process} in which we evaluate the transition from the last message that was classified as 'Benign' ($msg_{lastBen}$). If this transition is classified as 'Benign' than the classification of $msg_{i}$ will be updated to 'Benign'.   

\begin{algorithm}[th!]
	\caption{\label{alg:Detecting-order-timing-anomalies}Detect order and timing anomalies}
	\begin{algorithmic}[1]
		\Procedure{DetectAnomaly}{$msg_{i},model$}
		\State $score \gets model.Apply(msg_{i},msg_{i-1})$
		\If{$score \geq tr_{a}$}
		\State $label \gets 'Benign'$
		\Else
		\State $label\gets 'Anomaly'$
		\If{$IsAnomaly(msg_{i-1})$}
		\State $score \gets model.Apply(msg_{i},msg_{lastBen})$
		\If{$score \geq tr_{a}$}
		\State $label \gets 'Benign'$
		\EndIf
		\EndIf
		\EndIf
		
		\Return $label$
		\EndProcedure
	\end{algorithmic}
\end{algorithm}

\vspace{-\baselineskip}
\section{\label{sec:experiment-and-results}Experiments and Results}
In order to evaluate the proposed method we conducted two experiments.
For the first experiment we established an operational testbed which consists of real 1553 hardware, identical to that used in operational platforms. The testbed is capable of simulating DoS and spoofing attacks that utilize behavior and message manipulation methods, and implements the sequence-based anomaly detection module.

For the second experiment we performed off-line evaluation of the module on real 1553 system logs. These logs does not contain attack record, thus we only evaluated the method in terms of false alarms.

\subsection{\label{subsec:Emulator-Architecture}Testbed architecture}

\begin{figure*}[h!]
	\begin{center}
		\includegraphics[scale=0.36]{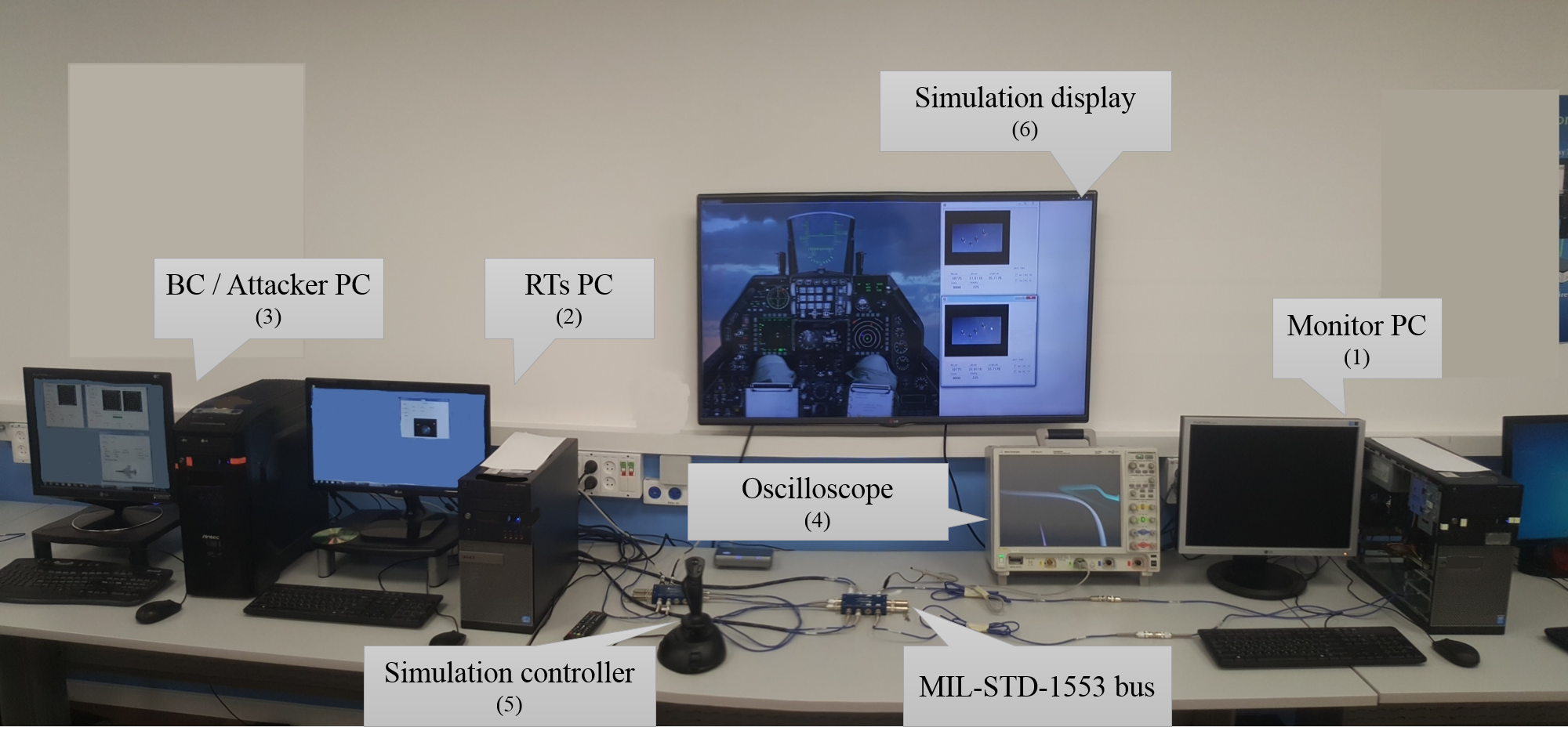}
		\caption{\label{fig:Simulator-architecture}Testing system architecture.}
		\vspace{-12pt}
	\end{center}
\end{figure*}

The testbed (illustrated in Figure \ref{fig:Simulator-architecture}) consists of three PCs that simulate the various components of simplified avionic systems. 
Each component is connected to the bus via a 1553 interface card. 
The subsystems are distributed on different PCs in order to physically place them at different distances from the bus, thereby simulating an actual bus topology setup. 
The main components simulated in the testbed are:

\textbf{BC.} This component initiates all communications over the bus. The BC is implemented in the BC/Attacker PC (item 3 in Figure \ref{fig:Simulator-architecture}) and is connected to the bus through a DDC BU-67114Hx interface. The BC's software includes a graphical user interface (GUI) that allows us to control its activity (legitimately or maliciously). 

\textbf{BM.} This component monitors the bus and implements our proposed sequence-based anomaly detection module.
The BM's software provides a GUI that enables online training of the module, online monitoring and detection of anomalous messages transmitted over the bus, and displaying visual alerts when such messages are detected. In addition, the BM writes to a log file 
all monitored messages and their labels (assigned by the anomaly detection algorithm) for offline analysis. The BM is implemented in 
the Monitor PC (item 1 in Figure \ref{fig:Simulator-architecture}) and is connected to the bus through a DDC BU-67114Hx interface.

\textbf{RTs.} The RTs' programs are implemented on the RTs PC
(item 2 in Figure \ref{fig:Simulator-architecture}). The RTs PC
is connected to the bus through two interfaces: a DDC BU-67114Hx interface and an Excalibur EXC-4000PCIe card which logically enables up to 32 connections. These RTs are used for benign activity simulation, and their software also provides a simple GUI which enables us to start or stop their operation.

\textbf{Attacker~component.} This component has the functionality of either a BC or an RT and is responsible for executing various attacks as a fake RT/BC (i.e., illegitimately connected to the bus) or a compromised RT/BC. The attacker component is controlled
manually through a GUI, and its software is implemented in
the same PC as the BC (item 3 in Figure \ref{fig:Simulator-architecture}) and is connected to the bus via another DDC BU-67114Hx interface.

The testbed also contains: (1) an oscilloscope (item 4 in Figure \ref{fig:Simulator-architecture}) for visualizing electric signals transmitted over the bus, (2) a controller (item 5 in Figure \ref{fig:Simulator-architecture}) for simulating user operations, and (3) a display (item 6 in Figure \ref{fig:Simulator-architecture}) for visualizing the physical impact on the simulated operations on the simulated system.

\textbf{Implemented attack scenarios.} In order to evaluate the proposed algorithm we implemented three attack scenarios: two spoofing scenarios and a DoS scenario that were
simulated on two different bus topologies. 
The topologies model two systems which operate differently from one another.
The first topology, denoted by $topology_1$, simulates a larger and more crowded system that consists of 19 components: the BC, BM, a compromised RT with BC capabilities (the attacker component), and 16 benign RTs. 
The major frame length is 80 milliseconds, during which 21 different messages are sent. Twenty of these messages have a time cycle of 20 milliseconds, and one has a time cycle of 80 milliseconds.
The second topology, denoted by $topology_2$, simulates a smaller system that performs some critical operation and consists of four components: the BC, BM, a compromised RT with BC capabilities (the attacker component), and a benign RT.
The major frame length is 20 milliseconds, during which five different messages are sent (the time cycle of each message is 20 milliseconds). 
Table \ref{tab:attck-scenarios-desc}
describes the normal and attack operations for each scenario, the topology it was implemented on, and the attack trigger.

\begin{table}[th!]
	\vspace{-\baselineskip}
	\centering \caption{\label{tab:attck-scenarios-desc}Attack scenario descriptions}
	\fontsize{6}{7}\selectfont
	{\renewcommand{\arraystretch}{1.5}
		\begin{tabular}{|>{\centering}m{0.043\textwidth}|>{\centering}m{0.048\textwidth}|>{\raggedright}m{0.10\textwidth}|>{\raggedright}m{0.045\textwidth}|>{\raggedright}m{0.12\textwidth}|}
			\hline 
			Attack scenario & Topology & \centering Normal operation  & \centering Attack trigger & \centering Attack operation\tabularnewline
			\hline 
			\hline
			Spoofing attack \#1 & \hspace{-6pt}$topology_1$ & 
			The BC sends a transmit command to one
			of the benign RTs, the RT responds with its data and then the BC transmits data to all other RTs.
			& Detection of idle time & \vspace{-6pt}
			The attacker component utilizes idle time on the bus to transmit fake data that overrides the real data sent earlier by the benign RT. \tabularnewline
			\hline
			
			Spoofing attack \#2 &
			\vspace{48pt}\hspace{-9pt}
			\multirow{2}{0.05\textwidth}{$topology_2$} &   
			\vspace{-57pt}
			\multirow{2}{0.11\textwidth}{The BC queries certain values from the two TRs, then broadcasts updates.
				When the user initiates the special operation, 
				the BC starts to broadcast different data and the RTs change their states accordingly. When the operation ends, the components return to their previous state and behavior (query and broadcast)} & 
			\multirow{2}{0.05\textwidth}{\vspace{15pt}\begin{center} Detection of user operation \end{center}} &
			When the compromised RT detects the start of the special operation, it pretends to be the BC and thwarts the operation by broadcasting fake data which causes the operation to fail. 
			\tabularnewline
			\cline{0-0} \cline{5-5}
			
			DoS attack &  &  &  &
			When the compromised RT detects the start of the special operation, it thwarts it by sending a large amount of fake commands to random RT addresses, causing collisions on the bus that lead to the failure of the  operation.\tabularnewline
			\hline 
	\end{tabular}}
\end{table}

\subsection{Testbed experiment}
In this experiment we evaluated the ability of our proposed method to identify the attack methods described in Table \ref{tab:attck-scenarios-desc}.

\textbf{Dataset~description.} The dataset used for this experiment was recorded by the BM in the testbed during both normal and attack activity. 
The dataset consists of three logs -- one for each attack scenario. A log is a collection of events (i.e., records), each representing a message. 
Each record contains all of the information extracted from the words (according to the word structure in Figure \ref{fig:Communication-protocol-words}). 
Each recording starts with a period of normal operation, and then the attack was activated and stopped via the attacker component's GUI. The anomalous messages were labeled accordingly by the attacker component.
Table \ref{tab:testbed-dataset-statistics} presents statistics about the dataset that was used for the evaluation. 

\begin{table}[th!]
	\caption{\label{tab:testbed-dataset-statistics}Testbed dataset statistics}
	\fontsize{6}{6}\selectfont
	\begin{center}
		{\renewcommand{\arraystretch}{1.3}
			\begin{tabular}{|>{\centering}m{0.11\textwidth}|>{\centering}m{0.06\textwidth}|>{\centering}m{0.06\textwidth}|>{\centering}m{0.06\textwidth}|>{\centering}m{0.06\textwidth}|}
				\hline 
				Attack scenario  & Train time period  & Train records \# & Test records \# & Anomaly \% \tabularnewline
				\hline 
				\hline 
				Spoofing attack \#1 & 4.5 sec. & 4,579 & 10,021 & 3.2\tabularnewline
				\hline 
				Spoofing attack \#2 & 4 sec. & 1,004 & 5,836 & 13\tabularnewline
				\hline 
				DoS attack & 4.2 sec. & 1,074 & 6,528 & 35.9\tabularnewline
				\hline 
		\end{tabular}}
	\end{center}
\end{table}

\textbf{Experiment~description.} The logs recorded by the BM were divided into two mutually exclusive sets (while preserving the chronological order of the records): (1) \emph{Training set - } a period of time from the beginning of the log (approximately four seconds) that contains only records of the system's normal operation (i.e., benign records); (2) \emph{Test set - } the rest of the log which contains both normal operation (benign) and attack (anomaly) records. 

The training sets were used to build a model representing the normal behavior of the system in each attack scenario, and the test sets were used to evaluate these representations.
After the training phase, each model was applied on the corresponding test set (as described in \ref{subsec:detection-process}). The labels ('Anomaly' or 'Benign') assigned by the detection algorithm were logged to a result file that was used for the performance evaluation of the model. 

\textbf{Results.} We evaluated the detection algorithm in terms of precision
and recall. The results are summarized in Table \ref{tab:testbed-exp-results}. As can be observed, the models learned during this experiment correctly identified all of the benign and anomalous messages (i.e., detected all anomalies with zero false alarms) in all of the attack scenarios, which is highly important for a mission critical system based on the MIL-STD-1553 communication bus.  

\begin{table}[h!]
	\caption{\label{tab:testbed-exp-results} Testbed experiments results}
	\fontsize{6}{6}\selectfont
	\centering
	{\renewcommand{\arraystretch}{1.5}
		\begin{tabular}{|>{\centering}m{0.09\textwidth}|>{\centering}m{0.04\textwidth}|>{\centering}m{0.03\textwidth}|>{\centering}m{0.04\textwidth}|>{\centering}m{0.03\textwidth}|>{\centering}m{0.04\textwidth}|>{\centering}m{0.03\textwidth}|}
			\hline 
			\multirow{2}{0.10\textwidth}{\centering Attack scenario} & \multicolumn{2}{c|}{Records \#} & \multicolumn{2}{c|}{Precision} & \multicolumn{2}{c|}{Recall}\tabularnewline
			\cline{2-7}
			& Anomaly & Benign & Anomaly & Benign & Anomaly & Benign\tabularnewline
			\hline 
			\vspace{1pt}
			Spoofing attack \#1  & 324 & 9,675 & 1.0 & 1.0 & 1.0 & 1.0\tabularnewline
			\hline 
			\vspace{1pt}
			Spoofing attack \#2  & 761 & 5,069 & 1.0 & 1.0 & 1.0 & 1.0\tabularnewline
			\hline 
			\vspace{1pt}
			DoS attack & 2,347 & 4,175 & 1.0 & 1.0 & 1.0 & 1.0\tabularnewline
			\hline 
	\end{tabular}}
\end{table}

\subsection{Real system experiment}
In this experiment we evaluated the performance of the proposed method on logs collected from a real 1553 system. Since these logs contain only the normal behavior of the monitored system, we used them in order to evaluate the minimal training time period required for achieving a good representation of the system.

\textbf{Dataset~description.} The dataset used for the second experiment was recorded in a real system using the MIL-STD-1553 data bus. This dataset consists of two logs recorded during different operations of the system and do not contain any abnormal behavior. The logs were formatted similarly to the testbed logs. Table \ref{tab:real-dataset-statistics} presents statistics of the collected datasets, the actual topology however, cannot be presented due to confidentiality issues.
The datasets were divided into chronologically ordered segments for incrementally training the module.
This dataset was used to estimate the minimal training  period for our method to correctly model a real system.

\begin{table}[th!]
	\caption{\label{tab:real-dataset-statistics}Real dataset statistics}
	\fontsize{6}{6}\selectfont
	\begin{center}
		{\renewcommand{\arraystretch}{1.3}
			\begin{tabular}{|>{\centering}m{0.03\textwidth}|>{\centering}m{0.075\textwidth}|>{\centering}m{0.07\textwidth}|>{\centering}m{0.05\textwidth}|>{\centering}m{0.05\textwidth}|>{\centering}m{0.06\textwidth}|}
				\hline 
				Log & Components \#  & Distinct messages \# & Duration & Records \# & Segments \#\tabularnewline
				\hline 
				$Log_{1}$ & 17 & 37 & 68.1 sec. & 30,034 & 56\tabularnewline
				\hline 
				$Log_{2}$ & 20 & 43 & 83 sec. & 47,526 & 123\tabularnewline
				\hline 
		\end{tabular}}
	\end{center}
\end{table}

\textbf{Experiment~description.} This experiment was conducted in several iterations. In the first iteration, the system was trained using the first segment and evaluated on the remaining segments. Next, in each iteration the module was trained with an additional segment and evaluated on the remaining ones. 

\begin{figure}[h!]
	\begin{center}
		\includegraphics[scale=0.4]{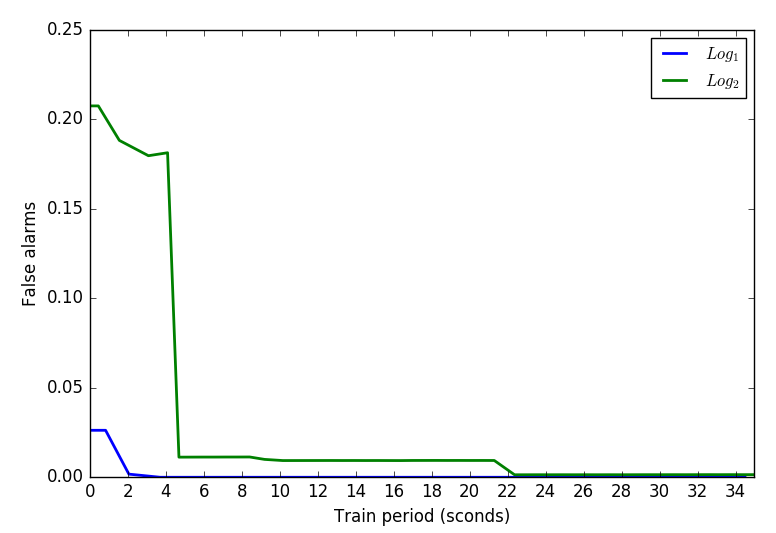}
		\caption{False alarm rate as a function of the training time period.}
		\label{fig:real-sys-exp-fa-vs-time}
	\end{center}
\end{figure}

\textbf{Results.} The module's performance was evaluated in terms of false alarm rate (i.e. benign messages that were labeled as anomalies). Figure \ref{fig:real-sys-exp-fa-vs-time} illustrates the experiment results for both logs. As can be observed, there is a very quick and significant decline in the false alarms rate for both logs. The false alarms rate dropped to zero after approximately three seconds for $Log_{1}$ and near zero after approximately five seconds for $Log_{2}$. Note that the false alarms rate for $Log_{2}$ further improved after about 17 more seconds.

\vspace{-2\baselineskip}
\section{\label{sec:discussion}Discussion}

As presented in the previous section, our method succeeded to distinguish between anomalous and benign messages in the testbed experiment and also learned and classified with very low false alarms rate the behavior of a real MIL-STD-1553 based system.

Although our method achieved very good detection rates, it is not capable of detecting all attack methods presented in Section \ref{subsec:threats}. Attack methods that utilize only data or status words, or require impersonating a component (i.e. spoofing, in this case, the threat agent sends benign messages with false data) will most likely not be detected by the suggested method, since it does not extract features from the data or status words, and does not have the ability to physically authenticate the components connected to the bus.

\begin{figure}[th!]
	\begin{centering}
		\includegraphics[scale=0.40]{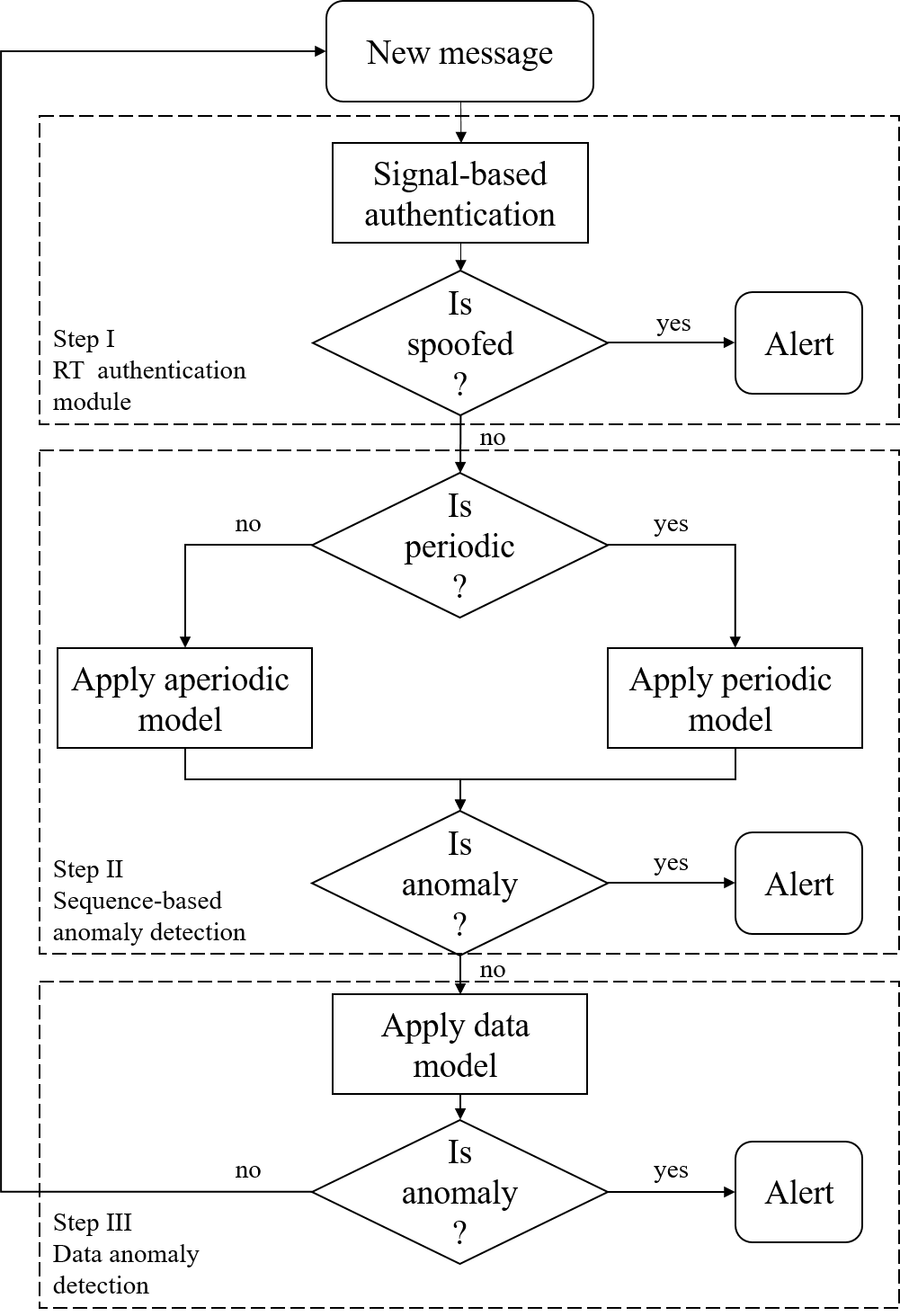}
		\par\end{centering}
		\caption{\label{fig:AnomDetectionProcess}Intrusion detection process.
	}
\end{figure}

Therefor, we would like to suggest an extended stepwise IDS architecture (Figure \ref{fig:AnomDetectionProcess}), which consists of three main conceptual detection modules: (1) signal-based RT authentication (Step I); (2) sequence and timing-based detection of anomalous messages (Step II); and (3) detection of data anomalies (Step III).

The detection process is designed as a cascading process that monitors the bus continuously and applies three detection modules.
As illustrated in Figure \ref{fig:AnomDetectionProcess}, the RT authentication module is the first module to be applied 
on the current message observed on the bus. 
If this module determines that the message is anomalous, meaning it was not sent by the expected component, an alert is generated.
Otherwise, the message passes on to the sequence anomaly detection module in order to verify that it was sent based on
the correct order and timing (as described in section \ref{sec:seq-timing-module}). If this module finds the message to be anomalous, indicating that the sender deviates from its normal behavior,
an alert is generated. Otherwise, the message passes on to the third and final module, the data anomaly module, which is used for detecting anomalies in the message's
payload. If no anomalies are detected in this step, the message is considered benign, and the IDS checks the next message.

\begin{figure}[th!]
	\begin{center}
		\includegraphics[scale=0.405]{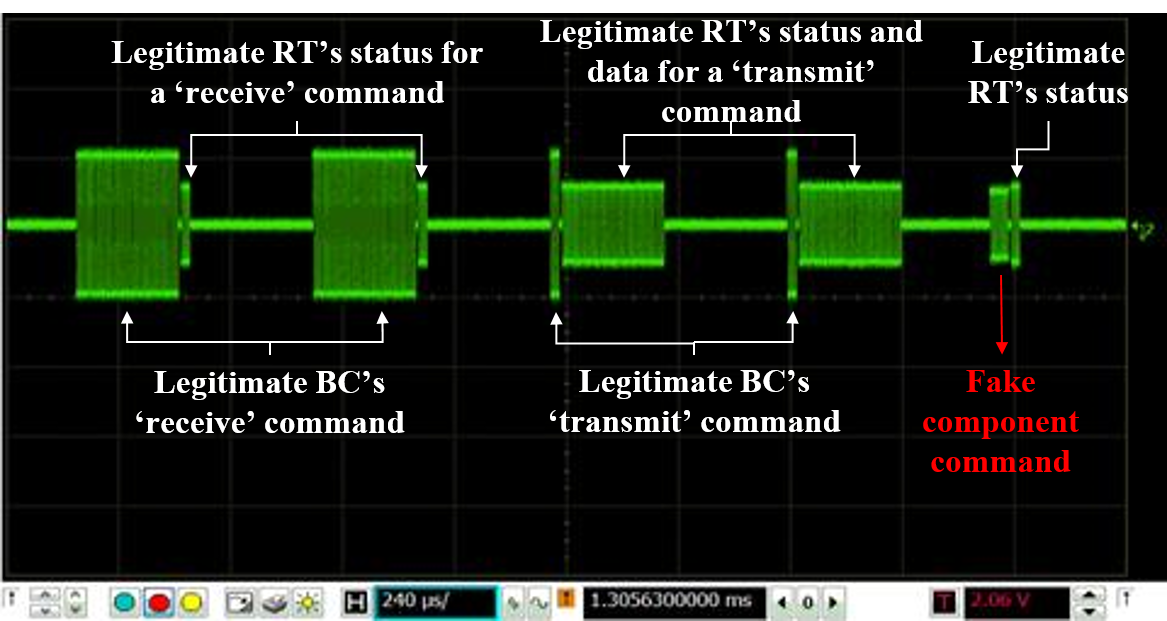}
		\caption{\label{fig:signal-sample}Signal samples captured by the testbed.}
	\end{center}
\end{figure}

\textbf{RT~authentication~module.}
Malicious nodes on the bus can generate spoofed messages and impersonate other legitimate RTs. Spoofed messages can be sent in their expected order and time, while carrying false payload, and thus will be difficult to detect. Therefore the first step in identifying anomalous messages should be authenticating the identity of the transmitting component.
We propose authenticating each RT by analyzing the electrical signal of the transmitting node as observed by the IDS. Various features that can be extracted from the electrical signals transmitted over the bus 
are correlated with the location of the transmitting component on the bus. Therefore, during the training/learning phase, given a dataset of legitimate communication captured by the IDS, it is possible to create a unique electrical profile for each transmitting component. During the operational (detection) phase, upon arrival of a new message, the IDS attempts to match the electrical signal patterns of each message with the profile of the transmitting component.
If no match is found, the IDS will issue an alert indicating that the message is spoofed and transmitted by a fake source.
Figure \ref{fig:signal-sample} is a capture of legitimate and fake signals from our testbed (described in Section \ref{subsec:Emulator-Architecture}).

\textbf{Data~anomaly~detection~module.}
This step of the detection process focuses on more sophisticated attacks involving invalidating 
the integrity of the data transmitted via the data words. 
A message that is analyzed by this module has successfully managed to bypass the previous two mechanisms. 
An attacker can manipulate the data words of benign messages (by either injecting false data or corrupting the data words) and disrupt the normal operation of the system.
This type of attack is easy to miss when focusing on just the command and timing features of messages, and it can have fatal consequences.

This module may contain more than one detection model, since different subsystems connected to the bus have different behavior and are using different data types (e.g., GPS location, speed, and status of different subsystems). Moreover, each data type may require the application of different machine learning algorithms to achieve optimal performance.

Data-related features that are extracted in order to apply the data anomaly detection module can be generic or application-specific. Generic features are computed from the raw bit stream data (e.g., edit distance computed on the current and previously transmitted data words, and byte distribution). Application-specific features are computed at a higher level of abstraction and can be correlated with other extracted information to detect anomalies. For example, a model analyzing location data can identify anomalies by detecting significant deviations from known routes, or detect suspicious behavior by correlating the location data with unlikely changes in speed, acceleration, and heading.
After extracting the features from the data words, an approach similar to \cite{kang2016DNN} can be applied for detecting anomalous data.

\section{\label{sec:future-work}Conclusions and Future Work}

In this paper we present a security analysis of MIL-STD-1553 
and suggested a machine learning-based approach as a possible solution for attacks detection. As a proof of concept we established
a testbed, evaluated the proposed sequence-based module on spoofing and DoS attacks simulated in the testbed and on logs recorded in a real system.
The results of the experiments showed that the suggested algorithm can distinguish between anomalous and legitimate messages with high level of accuracy and requires a very short period of training time to achieve a good representation of the system.

In future work we are planning to implement and evaluate additional attack scenarios, testing the proposed intrusion detection module on additional bus configurations, and perform additional evaluation on real systems data. Moreover, we plan to implement and evaluate the RT authentication module and the data anomaly detection module that will focus on detecting anomalies in geo-location data.

{\footnotesize{}\bibliographystyle{IEEEtran}
	\bibliography{Bibliography}

\begin{thebibliography}{10}

\bibitem{chandola2009anomaly}
{\sc Chandola, V., Banerjee, A., and Kumar, V.}
\newblock Anomaly detection: A survey.
\newblock {\em ACM computing surveys (CSUR) 41}, 3 (2009), 15.

\bibitem{checkoway2011comprehensive}
{\sc Checkoway, S., McCoy, D., Kantor, B., Anderson, D., Shacham, H., Savage,
  S., Koscher, K., Czeskis, A., Roesner, F., Kohno, T., et~al.}
\newblock Comprehensive experimental analyses of automotive attack surfaces.
\newblock In {\em USENIX Security Symposium\/} (2011), San Francisco.

\bibitem{cho2016fingerprinting}
{\sc Cho, K.-T., and Shin, K.~G.}
\newblock Fingerprinting electronic control units for vehicle intrusion
  detection.
\newblock In {\em 25th USENIX Security Symposium (USENIX Security 16)\/}
  (Austin, TX, 2016), USENIX Association, pp.~911--927.

\bibitem{chong2005Survivability}
{\sc Chong, J., Pal, P., Atigetchi, M., Rubel, P., and Webber, F.}
\newblock Survivability architecture of a mission critical system: the dpasa
  example.
\newblock In {\em 21st Annual Computer Security Applications Conference
  (ACSAC'05)\/} (Dec 2005), pp.~10 pp.--504.

\bibitem{deshu1991guilin}
{\sc Deshu, C., and Jixiang, W.}
\newblock Guilin institute of optical communications; fiber-optic mechanization
  of an aircraft internal time division command/response multiplex data bus
  (come up for discussions)[j].
\newblock {\em Optical Communication Technology Z 1\/} (1991).

\bibitem{editionmil}
{\sc EDITION, S.}
\newblock Mil-std-1553 designer's guide.

\bibitem{garcia2009anomaly}
{\sc Garcia-Teodoro, P., Diaz-Verdejo, J., Maci{\'a}-Fern{\'a}ndez, G., and
  V{\'a}zquez, E.}
\newblock Anomaly-based network intrusion detection: Techniques, systems and
  challenges.
\newblock {\em computers \& security 28}, 1 (2009), 18--28.

\bibitem{gillen1992introduction}
{\sc Gillen, A., and Shelton, J.}
\newblock Introduction of 3910 high speed data bus.
\newblock In {\em Military Communications Conference, 1992. MILCOM'92,
  Conference Record. Communications-Fusing Command, Control and Intelligence.,
  IEEE\/} (1992), IEEE, pp.~956--960.

\bibitem{gligor1983note}
{\sc Gligor, V.~D.}
\newblock A note on the denial-of-service problem.
\newblock In {\em IEEE Symposium on Security and Privacy\/} (1983),
  pp.~139--149.

\bibitem{hoppe2008security}
{\sc Hoppe, T., Kiltz, S., and Dittmann, J.}
\newblock {\em Security Threats to Automotive CAN Networks -- Practical
  Examples and Selected Short-Term Countermeasures}.
\newblock Springer Berlin Heidelberg, Berlin, Heidelberg, 2008, pp.~235--248.

\bibitem{roufa2010security}
{\sc Ishtiaq~Roufa, R.~M., Mustafaa, H., Travis~Taylora, S.~O., Xua, W.,
  Gruteserb, M., Trappeb, W., and Seskarb, I.}
\newblock Security and privacy vulnerabilities of in-car wireless networks: A
  tire pressure monitoring system case study.
\newblock In {\em 19th USENIX Security Symposium, Washington DC\/} (2010),
  pp.~11--13.

\bibitem{jiang2010periofic}
{\sc Jiang, W., Guo, W., and Sang, N.}
\newblock Periodic real-time message scheduling for confidentiality-aware
  cyber-physical system in wireless networks.
\newblock In {\em 2010 Fifth International Conference on Frontier of Computer
  Science and Technology\/} (Aug 2010), pp.~355--360.

\bibitem{kang2016DNN}
{\sc Kang, M.-J., and Kang, J.-W.}
\newblock Intrusion detection system using deep neural network for in-vehicle
  network security.
\newblock {\em PLOS ONE 11}, 6 (06 2016), 1--17.

\bibitem{klegerger2011security}
{\sc Kleberger, P., Olovsson, T., and Jonsson, E.}
\newblock Security aspects of the in-vehicle network in the connected car.
\newblock In {\em 2011 IEEE Intelligent Vehicles Symposium (IV)\/} (June 2011),
  pp.~528--533.

\bibitem{kuhn1998soft}
{\sc Kuhn, M.~G., and Anderson, R.~J.}
\newblock Soft tempest: Hidden data transmission using electromagnetic
  emanations.
\newblock In {\em International Workshop on Information Hiding\/} (1998),
  Springer, pp.~124--142.

\bibitem{lindsay2013stuxnet}
{\sc Lindsay, J.~R.}
\newblock Stuxnet and the limits of cyber warfare.
\newblock {\em Security Studies 22}, 3 (2013), 365--404.

\bibitem{liu2011false}
{\sc Liu, Y., Ning, P., and Reiter, M.~K.}
\newblock False data injection attacks against state estimation in electric
  power grids.
\newblock {\em ACM Transactions on Information and System Security (TISSEC)
  14}, 1 (2011), 13.

\bibitem{matsumoto2012preventing}
{\sc Matsumoto, T., Hata, M., Tanabe, M., Yoshioka, K., and Oishi, K.}
\newblock A method of preventing unauthorized data transmission in controller
  area network.
\newblock In {\em 2012 IEEE 75th Vehicular Technology Conference (VTC
  Spring)\/} (May 2012), pp.~1--5.

\bibitem{mcgraw2014cyber}
{\sc McGraw, R.~M., Fowler, M.~J., Umphress, D., and MacDonald, R.~A.}
\newblock Cyber threat impact assessment and analysis for space vehicle
  architectures.
\newblock In {\em SPIE Defense+ Security\/} (2014), International Society for
  Optics and Photonics, pp.~90850K--90850K.

\bibitem{miller2012scada}
{\sc Miller, B., and Rowe, D.}
\newblock A survey scada of and critical infrastructure incidents.
\newblock In {\em Proceedings of the 1st Annual Conference on Research in
  Information Technology\/} (2012), RIIT '12, ACM, pp.~51--56.

\bibitem{miller2015remote}
{\sc Miller, C., and Valasek, C.}
\newblock Remote exploitation of an unaltered passenger vehicle.
\newblock {\em Black Hat USA 2015\/} (2015).

\bibitem{mo2010false}
{\sc Mo, Y., and Sinopoli, B.}
\newblock False data injection attacks in control systems.
\newblock In {\em Preprints of the 1st workshop on Secure Control Systems\/}
  (2010), pp.~1--6.

\bibitem{muter2011entropy}
{\sc Müter, M., and Asaj, N.}
\newblock Entropy-based anomaly detection for in-vehicle networks.
\newblock In {\em 2011 IEEE Intelligent Vehicles Symposium (IV)\/} (June 2011),
  pp.~1110--1115.

\bibitem{nguyen2015towards}
{\sc Nguyen, T.~D.}
\newblock Towards mil-std-1553b covert channel analysis.
\newblock Tech. rep., Monterey, California. Naval Postgraduate School, 2015.

\bibitem{song2016intrusion}
{\sc Song, H.~M., Kim, H.~R., and Kim, H.~K.}
\newblock Intrusion detection system based on the analysis of time intervals of
  can messages for in-vehicle network.
\newblock In {\em 2016 International Conference on Information Networking
  (ICOIN)\/} (Jan 2016), pp.~63--68.

\bibitem{vai2016design}
{\sc Vai, M., Whelihan, D., Evancich, N., Kwak, K.~J., Li, J., Britton, M.,
  Foley, J., Lynch, M., Schafer, D., and DeMatteis, J.}
\newblock Systems design of cybersecurity in embedded systems.
\newblock In {\em 2016 IEEE High Performance Extreme Computing Conference
  (HPEC)\/} (Sept 2016), pp.~1--6.

\bibitem{wolf2004security}
{\sc Wolf, M., Weimerskirch, A., and Paar, C.}
\newblock Security in automotive bus systems.
\newblock In {\em Workshop on Embedded Security in Cars\/} (2004).

\bibitem{ye2000markov}
{\sc Ye, N., et~al.}
\newblock A markov chain model of temporal behavior for anomaly detection.
\newblock In {\em Proceedings of the 2000 IEEE Systems, Man, and Cybernetics
  Information Assurance and Security Workshop\/} (2000), vol.~166, West Point,
  NY, p.~169.

\end{thebibliography}
}{\footnotesize \par}

\end{document}